\documentclass{article}

\usepackage{arxiv}

\usepackage[utf8]{inputenc} 
\usepackage[T1]{fontenc}    
\usepackage{hyperref}       
\usepackage{xcolor}
\definecolor{darkblue}{rgb}{0, 0, 0.5}
\hypersetup{colorlinks=true, citecolor=darkblue, linkcolor=darkblue, urlcolor=darkblue}
\usepackage{url}            
\usepackage{booktabs}       
\usepackage{amsfonts}       
\usepackage{nicefrac}       
\usepackage{microtype}      
\usepackage{lipsum}		
\usepackage{graphicx}
\usepackage{natbib}
\usepackage{doi}

\usepackage{times,latexsym}
\usepackage{url}
\usepackage[T1]{fontenc}
\usepackage{graphicx}
\usepackage{subfigure}
\usepackage{tabularx}
\usepackage{etoolbox}
\usepackage{bbm}
\usepackage{multirow}
\usepackage{enumitem}
\usepackage{amssymb}
\usepackage{amsmath}
\usepackage{svg}
\usepackage{amsthm}
\usepackage{xspace}
\usepackage{float}
\usepackage{booktabs}       
\usepackage{amsfonts}       
\usepackage{nicefrac}       
\usepackage{color, soul}
\usepackage{setspace}
\usepackage{algorithm}
\usepackage{algpseudocode}

\renewcommand\cite{\citep}	

\renewcommand{\headeright}{}
\renewcommand{\undertitle}{}
\renewcommand{\shorttitle}{\textit{}}

\hypersetup{
pdftitle={Private QA},
pdfsubject={q-bio.NC, q-bio.QM},
pdfauthor={David S.~Hippocampus, Elias D.~Striatum},
pdfkeywords={First keyword, Second keyword, More},
}

\title{Reasoning over Public and Private Data in Retrieval-Based Systems}

\author{Simran Arora \\
  Stanford University \\
  Stanford, CA \\
  \texttt{simran@cs.stanford.edu}
  \\\And
  Patrick Lewis \\
  Facebook AI Research \\
  London \\
  \texttt{plewis@fb.com} \\\And
  Angela Fan \\
  Facebook AI Research \\
  Paris \\
  \texttt{angelafan@fb.com} \\\And
  Jacob Kahn* \\
  Facebook AI Research \\
  Menlo Park, CA \\
  \texttt{jacobkahn@fb.com} \\\And
  Christopher Ré* \\
  Stanford University \\
  Stanford, CA \\
  \texttt{chrismre@cs.stanford.edu}\thanks{Work done with equal contribution from Facebook AI Research and Stanford University.} \\
}

\newcommand{\blue}[1]{{\color{blue}#1}}

\newcommand{\datasetname}{\textsc{ConcurrentQA}\xspace}
\newcommand{\problemlongname}{\textsc{Public-Private Autoregressive Information Retrieval}\xspace}
\newcommand{\problemshortname}{\textsc{PAIR}\xspace}

\begin{document}
\maketitle
\begin{abstract}
Users and organizations are generating ever-increasing amounts of private data from a wide range of sources. Incorporating private data is important to personalize open-domain applications such as question-answering, fact-checking, and personal assistants. 
State-of-the-art systems for these tasks explicitly retrieve relevant information to a user question from a background corpus before producing an answer. While today's retrieval systems assume the corpus is fully accessible, users are often unable or unwilling to expose their private data to entities hosting public data.
We first define the \problemlongname (\problemshortname) privacy framework for the novel retrieval setting over multiple privacy scopes. We then argue that an adequate benchmark is missing to study \problemshortname since existing textual benchmarks require retrieving from a single data distribution. However, public and private data intuitively reflect different distributions, motivating us to create \datasetname, the first textual QA benchmark to require concurrent retrieval over multiple data-distributions.
Finally, we show that existing systems face large privacy vs. performance tradeoffs when applied to our proposed retrieval setting and investigate how to mitigate these tradeoffs.
\end{abstract}

\section{Introduction}

The world's information is split between that which is publicly and privately accessible and the ability to simultaneously reason over information from both scopes is useful to support personalized tasks. However, retrieval-based machine learning (ML) systems, which first collect relevant information to a user input from a background knowledge source before producing an output, do not consider retrieving from the private data that organizations and individuals aggregate locally. Retrieval systems are achieving impressive performance across open-domain applications such as language-modeling \cite{borgeaud2021retro},  question-answering \cite{voorhees1999trec,chen2017retrieveread}, and dialogue \cite{dinan2019wow}, and also benefit from practical properties such as updatability and a degree of interpretability. In this work, we focus on the underexplored question of how to personalize these systems while preserving privacy.

\begin{figure}
    \centering
    \includegraphics[width=0.95\linewidth]{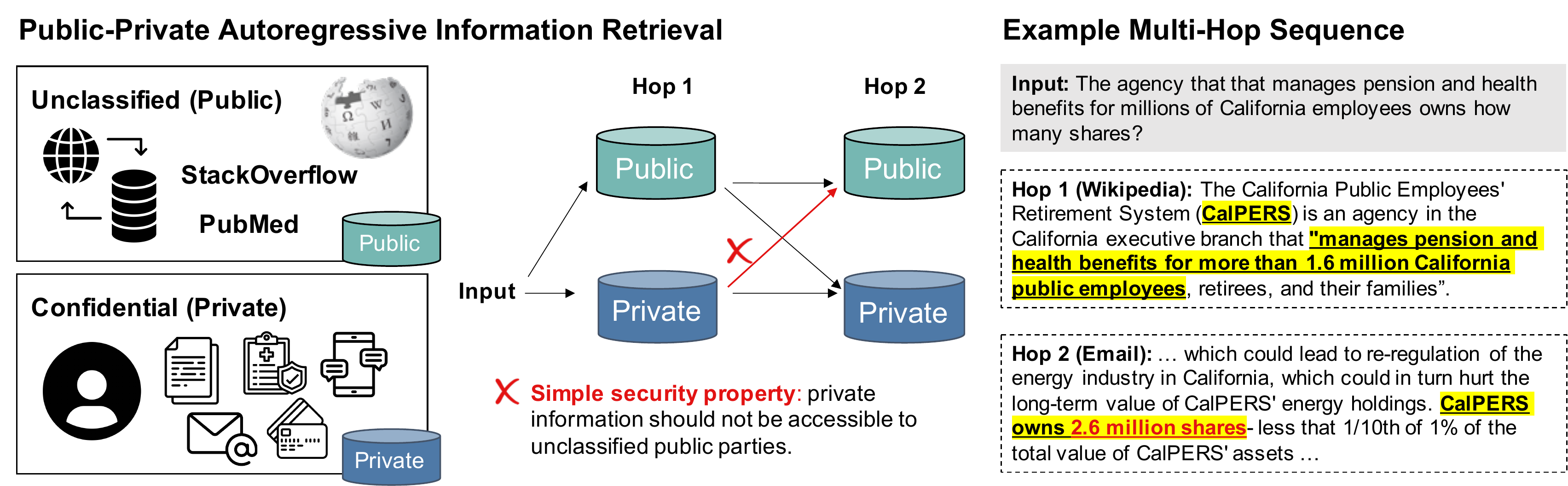}
    \caption[width=\linewidth]{Multi-hop retrieval systems use beam search to retrieve data from a background corpus: a document retrieved in round, or \textit{$\mathrm{hop_i}$}, is used to retrieve in $\mathrm{hop_{i+1}}$. Thus, if private documents (e.g., medical records or emails) were retrieved in $\mathrm{hop_i}$, it would sacrifice privacy if they were used to retrieve public information in $\mathrm{hop_{i+1}}$, since the document would be exposed to the entity hosting the public data. Existing multi-hop systems do not consider retrieval from multiple privacy scopes, the focus of this work.}
    \label{fig:main}
    \vspace{-5mm}
\end{figure}

Consider the following examples that require a combination of public and \blue{private} information: individuals could ask \textit{``With \blue{my GPA and SAT score}, which universities should I apply to in the United States?''} or \textit{``Is \blue{my blood pressure} in the normal range for someone 55+?''}. In an organization, an ML engineer could ask: \textit{``How do I fine-tune a language model, based on public StackOverflow and \blue{our internal company documentation}?''},\footnote{\url{https://stackoverflow.com/}} or a doctor could ask \textit{``How are COVID-19 vaccinations affecting patients with type-1 diabetes based on \blue{our private hospital records} and public PubMed reports?''}.\footnote{\url{https://pubmed.ncbi.nlm.nih.gov/}}
Currently, to answer such questions, users must manually cross-reference public and private information. A public knowledge source would not contain a user's medical record and a private knowledge source is not likely to include all medical statistics.
In this work, we propose a framework for using public (or global) information to enhance our understanding of private (or local) information, which we refer to as \textit{\problemlongname} (\problemshortname).

Given a user question, retrieval-based systems operate by collecting the most similar documents to the question from a massive corpus, and providing these to a separate model which can reason over the information to produce an answer \cite{chen2017retrieveread}. Answering complex questions about public and private data requires reasoning over information that is distributed across multiple documents (e.g., public medical statistics and private health records), termed multi-hop reasoning \cite{welbl2018mhtask}. For popular benchmarks of multi-hop questions, we find that introducing multiple rounds of retrieval, where the initial question \textit{combined with the text of documents} retrieved in round $i$ is used to retrieve in round $i+1$, provides upwards of $75\%$ performance gains versus using a single round of retrieval (Section \ref{sec:method}). Accordingly iterative retrieval is the typical approach for multi-hop reasoning \cite{miller2016kvm, feldman2019multihop, asai2020graphpathsqa, xiong2021mdr, qi2021irrrqa, khattab2021baleen}. 

Existing multi-hop systems assume retrieval is performed over one corpus, in a single privacy scope. However, data is often distributed across multiple parties with different privacy restrictions and in certain (e.g., government and medical) settings, data cannot be shared publicly. Broadly, users and organizations often do not want to expose all data to public entities, and it is unlikely that private parties can locally host terabyte-scale and constantly updating web data, naturally resulting in \textit{multiple corpora} over which to retrieve.

\textbf{Example} To understand why multi-hop retrieval over distributed corpora implicates different privacy concerns, consider two questions 
from an employee standpoint. First, `\textit{Of the products our top competitor released this month, which are most similar to our unannounced upcoming products?}''. To answer this question an existing multi-hop system likely (1) retrieves documents (e.g., news articles) about competitor releases from the public corpus, and (2) uses these to collect private documents (e.g., company emails and announcements) that detail upcoming products, so no private information is leaked. Meanwhile, ``\textit{Have any companies ever released similar products to the one we are designing?}'' entails retrieving (1) private documents about the upcoming product, and (2) \textit{using the confidential product design documents} to retrieve documents about public products. The latter retrieval reveals private company documents to the untrustworthy entity hosting the public corpus. An effective privacy model will preclude this private-then-public retrieval, preventing any possibility of leakage.

Guided by the constraint that in many situations users cannot or do not want to publicly expose their private information to public entities, we propose \problemshortname as a natural and effective privacy framework for complex QA. \problemshortname employs the classical Bell-LaPadula Model (BLP) (see Section \ref{sec:method} for details), a simple and efficient framework which guarantees no leakage of private data \cite{bell2976blm}. The framework was originally developed for, and widely used by, government agencies to successfully manage multi-security-level access control,
which maps to our setting of open-access public and private user data.

\paragraph{Study and evaluation with \problemshortname} 
We next address how to methodologically study and evaluate retrieval in the \problemshortname setting.
We first propose an adaptation of one of the most popular multi-hop benchmarks, HotpotQA \cite{yang2018hotpotqa}, which requires retrieval from Wikipedia data --- though we show this adaption is limited insofar as private and public documents are likely to come from different distributions.\footnote{For example, a Wikipedia passage focuses on a single entity, while private emails can cover many different topics in the same document, and information about a given entity can appear in many different emails.} We observe that \textit{all} existing textual multi-hop benchmarks require retrieving from a single domain such as Wiki or Freebase.
Thus, to more appropriately evaluate \problemshortname,
we create and release the first textual multi-domain, multi-hop benchmark, called \datasetname, which spans Wikipedia in the public domain and an open source email collection in the private domain.

Finally, we implement a \problemshortname-preserving retrieval system which excitingly answers many questions spanning public and private data. However, we show multi-hop reasoning systems exhibit high sensitivity to document retrieval order, thus presenting a privacy-performance tradeoff. Models sacrifice upwards of 19\% performance under \problemshortname constraints to protect document privacy and 57\% under constraints to protect query privacy, when compared to a baseline system with standard, non-privacy aware retrieval mechanics. 

In summary, we ask how to improve personalization while preserving privacy in retrieval-based systems and our particular contributions are:

\begin{itemize}
    \item We define the problem of retrieving over public and private data, and introduce the \textit{\problemlongname} (\problemshortname) privacy framework based on the classical Bell-LaPadula Model.
    \item We create \textit{\datasetname}, the first textual multi-domain, multi-hop benchmark. In the absence of privacy concerns, the benchmark allows studying  multi-distribution retrieval in general.
    \item  We demonstrate and quantify the privacy-performance tradeoff faced by existing multi-hop systems in \problemshortname and investigate challenges in mitigating the tradeoff. 
\end{itemize}

We hope the framework, resources, and analysis we present encourage further research towards building privacy-preserving retrieval systems.\footnote{We release all code and datasets: \url{https://github.com/facebookresearch/concurrentqa}}

\section{Background \& Related Work}

\label{sec:background}
\paragraph{Retrieval-Based Systems}
Open-domain applications in NLP, such as open-domain QA \cite{voorhees1999trec,chen2017retrieveread}, personal assistants \cite{dinan2019wow}, and language modeling \cite{borgeaud2021retro} need to support inputs across a broad range of topics.
\textit{Implicit-memory} approaches for open-domain tasks focus on memorizing knowledge within model parameters, for example by taking a pretrained language model such as T5 or BART and fine-tuning it on question-answer training pairs \cite{roberts2020t5}. 
In contrast, open-domain systems typically have access to the information in a background corpus (e.g., Wikipedia) or knowledge graph (e.g., Wikidata). Systems which explicitly exploit this information, called \textit{retrieval-based systems}, introduce a step to retrieve information that is relevant to the input from the background corpus, and provide this to a separate task model that produces the output. Retrieval-free approaches have not been shown to work convincingly in multi-hop settings \cite{xiong2021mdr}.

\paragraph{Multi-hop Open-Domain Question Answering}
We concretely demonstrate the challenges in applying \problemshortname to existing systems by focusing on open-domain QA (ODQA), a classic application for retrieval-based systems. ODQA entails providing an answer $a$ to a  question $q$, expressed in natural language and without explicitly provided context from which to find the answer \cite{voorhees1999trec}. Prevailing methods for ODQA collect a large collection of documents $D$ and follow a retrieve-and-read approach \cite[][inter alia.]{chen2017retrieveread} where the retriever retrieves a small set of relevant documents from the collection, from which the reader model extracts an answer. The answer is typically a span from one or more of the retrieved passages.

Our setting is concerned with complex queries where the supporting evidence for the answer is distributed across multiple (public and private) documents, termed multi-hop reasoning \cite{welbl2018mhtask}.
To collect the distributed evidence, existing multi-hop systems use multiple iterations of retrieval: representations of the passages retrieved in iteration $i$ are used to retrieve passages in iteration $i+1$. The beam search is a consistent backbone of multi-hop systems \cite{miller2016kvm, feldman2019multihop, asai2020graphpathsqa, wolfson2020break, xiong2021mdr, qi2021irrrqa, khattab2021baleen}. Various datasets have been developed for multi-hop reasoning \cite{yang2018hotpotqa, talmor2018complexwebqa}. We discuss the applicability of these benchmarks to the \problemshortname setting in Section \ref{sec:benchmark}. 

\paragraph{Privacy Framework} The proposed privacy framework, \problemshortname, is designed after the classical Bell-LaPaluda (BLP) privacy model  \cite{bell2976blm}, which has been widely and successfully used to manage access control between individuals of given clearance levels to objects of given classification levels. Our work instantiates BLP in the context of retrieval-based systems. Broadly, using freely available public resources, such as large models trained on public data and raw public data, locally, is a compelling set up because this paradigm incurs \textit{no privacy leakage} whatsoever and can inject personal knowledge without requiring training. This is in contrast to the Federated Learning (FL) \cite{mcmahan2016fl} and Differential Privacy (DP) \cite{dwork2006dp} privacy frameworks, which do leak information \cite{shokri2017meminference, nasr2019meminference}. Our setting resembles the FL setting in so far as data heterogeneity and distribution are properties of both, though FL has focused on collective model training across multiple parties, while we focus on information retrieval for a single individual or organization that owns private information. 
Overall, access control frameworks have been widely used in practice for many years \cite{hu2006nist}.

Proposed cryptographic methods for retrieval privacy include obfuscating the query by interleaving real and fake queries \cite{gervais2014queryobfusc} or performing
secure approximate nearest neighbor (ANN) search.
Applications such as search and personal assistants require low latency, and especially for complex queries requiring \textit{multiple hops} (i.e., the number of queries grows exponentially with the number of hops) over high dimensional vectors, the computational overhead of existing cryptographic methods is prohibitive \cite{zuber2021privatenn, chen2019privatennsanns}. Even secure ANN approaches that sacrifice some privacy leakage for better efficiency, are too slow for our setting \cite{schreiber2021PrivateNN}.
Other works propose fully on-device search engines, but scaling the amount of public data that can be hosted locally, not to mention updating at the of rate public data updates, remains challenging \cite{cao2019deqa}.

Ultimately, \problemshortname is a natural starting point in a rich decision space; we hope the resources we present facilitate research on alternate privacy models for public-private ODQA under differing cost-landscapes and privacy tradeoffs.

\section{Privacy-Aware ODQA Framework}
\label{sec:method}
This section presents our novel retrieval setting and \problemlongname privacy framework.

\subsection{Preliminaries}
\paragraph{Objective} Given a multi-hop input $q$, an individual or organization's private documents $p \in D_P$, and public documents $d \in D_G$, the objective is to provide the user with the correct answer $a$, which is a span in one or more of the documents. Figure \ref{fig:main} (Right) provides a multi-hop reasoning example, and instances of private and public data (Left).

\paragraph{Standard, Non-Privacy Aware QA} Standard non-private multi-hop ODQA involves answering $q$ with the help of passages $d \in D_G$, using beam search. In the first iteration of retrieval, the $k$ passages from the corpus, $d_1, ..., d_k$, that are most relevant to $q$ are retrieved. The text of a retrieved passage is combined with $q$ using a combination function $f$ (e.g., concatenating the query and passages sequences) to produce $q_i = f(q, d_i)$, for $i \in [1..k]$. Each $q_i$, which contains an explicitly retrieved document, is used to retrieve $k$ more passages in the following iteration.

\textit{Are multiple hops useful?} An important question is whether multiple-hops are actually required for answering complex questions. Differently from \citet{min2019compmultihop} and \citet{chen2019multihopdesign}, we consider this question in the open-domain setting. We observe that the performance on HotpotQA \cite{yang2018hotpotqa} improves by 26.4 EM (75\%) when using two iterations versus using one iteration (Appendix \ref{sec:appendix}).

\paragraph{Bell-LaPadula Model} The Bell-LaPadula Model (BLP) manages the access of \textit{subjects} with assigned clearance levels to \textit{objects} of assigned security levels \cite{bell2976blm}. BLP is defined by three security rules: subjects cannot read data at higher security levels (Simple Security Property), subjects cannot write to data-stores at lower security levels (*-Property), and discretionary access to objects can be granted or revoked from subjects (Discretionary Security Property). We next present our privacy framework based on BLP.

\subsection{\problemlongname Framework} In the private QA setting, users and organizations are classified and hold confidential private data, and unclassified services (e.g., cloud services) host public data. The user inputs to the public-private QA system are ${D_P}$ and ${q}$. We now describe the \problemshortname framework and challenges in applying non-private retrieval methods to both ${D_P}$ and ${D_G}$.

\paragraph{Constraint 1: Data is stored in two separate enclaves and personal documents $p \in \bf{D_P}$ can not leave the user's enclave.}  \problemshortname requires introducing a second, private corpus over which to retrieve, since users do not want to publicly expose their data to create a single public corpus nor blindly write personal data to a public location.\footnote{Following from the \textit{Simple Security Property} and \textit{*-Property} in the BLP model.} Further, we assume it is infeasible to copy public data to produce a single local corpus for each user. This is because not only are there terabytes of public data, but public data is also constantly being updated. Thus, users host a  private data (${D_P}$) and public (cloud) entities host open-access public data (${D_G}$).

Now given an input query $q$, the system must perform one retrieval over ${D_G}$ and a second over ${D_P}$. The top-$k$ retrieved passages for each iteration will include $k_P$ private passages and $k_G$ public passages the top $k$ of the $k_P + k_G$ passages are used for the following iteration of retrieval. 

If the retrieval-system stops after a \textbf{single-hop}, there is no privacy risk since no $p \in D_P$ is seen by public entities. \footnote{Single-hop can also avoid performance degradations arising from using two enclaves. Recall that a non-private system retrieves the top $k$ overall passages, so if for example $k_P = \frac{k}{2}$ and $k_G = \frac{k}{2}$, such that $k_P + k_G = k$, the system may not retrieve the optimal $k$ passages that the non-private system would have retrieved (e.g., consider when the overall top $k$ passages for a question are in $D_G$). However letting $k_P \in [0 .. k]$,  $k_G \in [0 .. k]$ circumvents this challenge, at the cost of retrieving a few more passages per hop.}
However for \textbf{multi-hop} questions, if $k_P > 0$ for an initial round of retrieval, meaning there exists some $p_i \in {D_P}$ which was in the top-$k$ passages, in general it would sacrifice privacy if $f(q, p_i)$ were to be used to perform the next round of retrieval from $D_G$. Thus to preserve the privacy of private documents, under \problemshortname, public retrievals precede private document retrievals.

\paragraph{Constraint 2: Inputs that entirely rely on private information should not be revealed publicly.} Given the two indices for ${D_P}$ and ${D_G}$, $q$ may be entirely answerable using multiple hops over the ${D_P}$ index, in which case, $q$ would never need to leave the user device. For example, consider the hypothetical query from an employee standpoint: \textit{Does the search team use any infrastructure tools that our personal assistant team does not use?}, which is answerable purely through private company information.
Prior work demonstrates that queries are very revealing of user interests, intents, and backgrounds \cite{xu2007personalsearch, gervais2014queryobfusc, hill2012target}, and for users who are especially concerned about privacy, there is an observable difference in their search behavior \cite{zimmerman2019querybehavior}.

Adherence to the \problemshortname framework allows no possibility for data leakage and is simple to understand, without introducing inefficiencies over the non-private baselines. The framework is, however, conservative, which, as we shall see, can have performance implications . 
If a user does not mind revealing certain data or weakening these constraints, our approach can be extended with methods that manage such user-specified exceptions \cite{xu2007personalsearch, shou2014personalsearch}. \problemshortname is a natural privacy framework, based on a widely used and successful foundation, BLP, however we hope this work inspires broader research on privacy-preserving solutions under alternate performance-privacy cost models.

\section{\datasetname for Multi-Domain Multi-Hop Reasoning}
\label{sec:benchmark}
In this section, we develop a testbed for studying the \problemshortname framework. The key requirement is a set of questions spanning two corpora, ${D_P}$ and ${D_G}$. We begin by considering the use of existing benchmarks and describing the limitations we encounter, motivating the creation of our new benchmark, \datasetname. Then we describe the benchmark collection process and provide an analysis of the contents.

\subsection{Adapting Existing Benchmarks to Privacy-Preserving QA and Limitations}  We first adapt the widely used benchmark, HotpotQA \cite{yang2018hotpotqa}, to study our problem. 
HotpotQA contains multi-hop questions, which are each answerable by multiple Wikipedia passages. We create HotpotQA-\problemshortname by splitting the Wikipedia corpus into ${D_G}$ and ${D_P}$ by randomly assigning Wikipedia articles to one or the other. This results in questions entirely reliant on $p \in {D_P}$, entirely reliant on $d \in {D_G}$, or reliant on a mix of one private and one public document, allowing us to evaluate performance under the \problemshortname constraints.

Ultimately however, $D_P$ and $D_G$ come from a single Wikipedia distribution in HopotQA-PAIR.
While it is possible that public and private data come from the same distribution (e.g., organizations routinely develop internal Wikis in the style of public Wikipedia), private and public data will intuitively often reflect different linguistic styles, structures, and topics, that further evolve over time \cite{hawking2004enterprisesearch}. 
We observe all existing textual multi-hop benchmarks focus on retrieving from a single distribution (Table \ref{tab:mutlihopqa}). Additionally, we cannot combine existing benchmarks over two different corpora because this will not yield questions requiring one passage from each domain. 
Methodologically, in the \problemshortname setting we likely will not have access to training data from all downstream (private) domains. 
To evaluate with a realistically private set of information and \problemshortname set up, we create a new benchmark \datasetname.

\begin{table}[t]
\begin{center}
\begin{tabular}{p{6cm}p{0.75cm}p{2.75cm}}
\toprule
Dataset & Size & Domain \\
\midrule
WebQuestions \cite{berant2013webq} & 6.6K & Freebase \\
WebQSP   \cite{yih2016webqsp}   & 4.7K  & Freebase \\
WebQComplex \cite{talmor2018complexwebqa} & 34K   & Freebase \\
MuSiQue  \cite{trivedi2021musique}   & 25K & Wiki \\
DROP \cite{dua2019drop}& 96K          & Wiki \\
HotpotQA  \cite{yang2018hotpotqa} & 112K     & Wiki \\
2Wiki2MultiHopQA \cite{ho20202wikidataset} & 193K & Wiki \\
Natural-QA  \cite{47761} &  300K & Wiki \\
\midrule
\datasetname &   18.4K  & Email \& Wiki \\
\bottomrule
\end{tabular}
\caption{Existing textual multi-hop benchmarks are designed over a single-domain.}
    \label{tab:mutlihopqa}
\end{center}
\end{table}

\begin{table*}[t]
\small
\begin{tabular}{p{5.3cm}p{9.8cm}}
\toprule
Question & Hop 1 and Hop 2 Gold Passages  \\
\midrule
What was the estimated 2016 population of the city that generates power at the Hetch Hetchy hydroelectric dams? &
\textit{Hop 1} An email mentions that San Francisco generates power at the Hetch Hetchy dams. \newline \textit{Hop 2} The Wikipedia passage about San Francisco reports the 2016 census-estimated population.
\\
\midrule
Which firm invested in both the fifth rounding of funding for Extraprise and first round of funding for JobsOnline.com? & 
\textit{Hop 1} An email reports the list of investors in the fifth round for Exraprise.\newline \textit{Hop 2} An email reports the list of investors in the first round for JobsOnline.com. 
\\
\midrule
What is the position of the person who sent an e-mail on 3/15/01 at 3:26 PM where the first listed recipient was Susan McCabe? &
\textit{Hop 1} An email that forwards an original email sent by Julee Malinowski-Ball.\newline
\textit{Hop 2} A different email from Julee Malinowski-Ball, which includes her position in the signature. \\
\midrule
The paper that ran a story on 4/20/01 titled "Hines will add to skyline" bought out its long-time rival in what year to become its home city's primary newspaper? &
\textit{Hop 1} An email includes a list of headlines, relevant to Enron, published by newspapers from 4/20/01. The article of interest was by the Houston Chronicle. \newline
\textit{Hop 2} The Wikipedia passage about the Houston Chronicle describes the 1995 buy-out of the rival. \\
\bottomrule
\end{tabular}
\caption{Example queries constructed over Wikipedia (${D_G}$) and emails (${D_P}$). }
    \label{tab:enron_demo_examples}
\end{table*}

\subsection{\datasetname Overview} 
We create and release a new multi-hop QA dataset, \textit{\datasetname}, which is designed to more closely resemble a practical use case for \problemshortname. \textit{\datasetname} contains 
questions spanning Wikipedia documents as $D_G$ and Enron employee emails \cite{klimt2004enron} as $D_P$. The email corpus is one of the only collections of real emails that has been publicly released for research use. \footnote{The Enron Corpus includes emails generated by 158 employees of Enron Corporation in the years before the company's collapse in 2001 due to accounting fraud. The corpus was generated from Enron email servers by the Federal Energy Regulatory Commission (FERC) during its investigation of the company.} We imagine two evaluation settings for \datasetname: (1) performance under defined (either \problemshortname or future proposals) privacy restrictions (presented in Section \ref{sec:solution}), and (2) 
multi-domain question-answering in the absence of privacy concerns
(presented in Section \ref{sec:section6}).

\paragraph{Contents} The full set of information collected from the crowd worker includes: the \textit{question} which requires reasoning over multiple documents, the \textit{answer} to the question which is a span in one of the documents, and the specific \textit{supporting sentences} in the documents which are necessary to arrive at the answer and can serve as useful supervision signals. Given an input question from our dataset, the QA system must extract a span of text from the contexts as the answer. 

\paragraph{Ethics Statement} The Enron Email Dataset is already widely-used in NLP research \cite{heller2017enronnyt}. That said, we acknowledge the origin of this data as collected and made public by the U.S. Federal Energy Regulatory Commission during their investigation of Enron. We note that many of the individuals whose emails appear in the dataset were not involved in any wrongdoing. We defer to using inboxes that are frequently used and well-studied in prior literature and that were not subject to redaction requests from affected employees, remaining freely-available in the public domain.

\subsection{Benchmark Design} 
As in HotpotQA, \datasetname is collected by showing crowd workers multiple supporting context documents and asking them to submit a question that requires reasoning over all the documents. 
We discuss the tradeoffs of our design choices in Section 4.5.

\paragraph{Passage Pairs}  As discussed in \citet{yang2018hotpotqa}, collecting a high-quality multi-hop QA dataset is challenging because it is important to provide \textit{reasonable} pairs of supporting context documents to the worker --- not all article pairs are conducive to a good multi-hop question. There are four types of pairs we need to collect for the $\mathrm{Hop_1}$ and $\mathrm{Hop_2}$ passages: Private and Private, Private and Public, Public and Private, and Public and Public.  We use the insight that we can obtain meaningful passage-pairs by showing workers passages that mention similar or overlapping entities. All crowdworker assignments contain unique passage pairs. We release all our code for creating the passage pairs from raw data and Algorithm \ref{CHalgorithm} gives the full data collection procedure.

While entity-tags are readily available for Wikipedia passages, hyperlinks are not readily available for many unstructured data sources including emails. Personal data also contains both private and public (e.g., Wiki) entities. High precision entity-linking is critical for the quality of the benchmark: for evaluation purposes, a question assumed to require the retrieval of private passages, should not be unknowingly answerable by public passages. We use a combination of off-the-shelf entity recognition and linking tools, and post-processing to tag private emails (Additional details in Appendix \ref{sec:appendix}). \footnote{\url{https://spacy.io/}, \url{https://github.com/egerber/spaCy-entity-linker}} For all passage-pairs shown to crowdworkers, we provide a hint that describes overlapping entities between the passages to assist with question generation.

\begin{table}[t]
\begin{center}
\begin{tabular}{p{1cm}p{1.5cm}p{2cm}p{1.5cm}}
\toprule
Split & Total & Comparison & Bridge \\
\midrule
Train & 15,239  &  1093 & 14,146 \\
Dev   & 1,600 & 200 & 1,400 \\
Test  & 1,600 & 200 & 1,400 \\
\bottomrule
\end{tabular}
\caption{\datasetname Benchmark size statistics. The evaluation sets are balanced between questions for which the gold evidence passages are emails versus Wikipedia passages for $\mathrm{Hop_1}$ and $\mathrm{Hop_2}$ respectively.}
\vspace{-6mm}
\label{tab:mutlihopqa_splits}
\end{center}
\end{table}

\paragraph{Dataset Collection}
The dataset collection proceeded in two stages, question generation and validation. 
Tasks were conducted through Amazon Mechanical Turk\footnote{\url{https://www.mturk.com/}} using the Mephisto interface.\footnote{\url{https://github.com/facebookresearch/Mephisto}} The end-to-end pipeline is in Figure \ref{fig:data_collection}. 

The question generation stage began with an onboarding process in which we provided training videos, documents with examples and explanations, and a multiple-choice exam. Workers completing the onboarding phase were given access to pilot assignments, which we manually reviewed to identify individuals providing high-quality submissions. Finally we worked with the shortlisted individuals to collect the full dataset. 

For validation, we manually reviewed over 2.5k queries to identify workers with high-quality submissions, and prioritized including manually-verified examples in the final test and dev splits. Through reviewing, we identified the key reasons to invalidate and exclude questions from the benchmark (e.g., if a question could be answered using one passage alone, has multiple plausible answers either in or out of the shown passages, or simply lacks clarity). Using these insights, we developed a second task to validate all generated queries. 
The validation task again involved onboarding and pilot steps, in which we manually reviewed performance. We shortlisted $\sim$20 crowdworkers with high quality submissions who collectively validated examples appearing in the final benchmark.

\subsection{Benchmark Analysis}
In this section we analyze the contents of \datasetname. The background corpora 
contain 47k email passages ($D_P$) and 5.2M Wikipedia passages ($D_G$), and the benchmark contains 18,439 total examples 
(Table \ref{tab:mutlihopqa_splits}). Table \ref{tab:enron_demo_examples} includes examples of \textit{\datasetname} queries.

\paragraph{Question Types} We identify three main reasoning patterns required for \datasetname questions: (1) \textit{bridge questions} require identifying an entity or fact in hop 1 on which the second retrieval is dependent, (2) \textit{attribute questions} require identifying the entity that satisfies all attributes in the question, where attributes are distributed across multiple passages, and (3) \textit{comparison questions} require comparing two similar entities, where each entity appears in a separate passage. We estimate the benchmark contains 80\% bridge, 12\% attribute, and 8\% comparison questions.

Salient topic categories that may be more popular in \datasetname compared to purely Wikipedia-based benchmarks include 
questions 
related to investments in projects and companies, 
newspaper articles,
executives or changes in company leadership (e.g., new board members, C-Suite),
legal activity (e.g., introduction, voting, or opposition to proposed bills or lawsuits and court cases),
email features (see Example 4 in Table \ref{tab:enron_demo_examples}), and political events.

\paragraph{Passage Types} There is a distinct shift between the emails and Wikipedia passages. \textbf{Format:} Wikipedia passages for entities of the same type tend to be similarly structured, while Enron emails introduce many formats --- for example, certain emails contain portions of forwarded emails, lists of articles, or spam advertisements. \textbf{Noise:} Wikipedia passages tend to be typo-free, while the emails contain several typos, URLs,  and inconsistent capitalization (examples in Table \ref{tab:additional_examples}). \textbf{Entity Distributions:} Wikipedia passages tend to focus on details about one entity, while a single email can cover multiple (possibly unrelated) topics. Information about Enron entities is also observationally more distributed across passages, whereas public entity-information tends to be localized to one Wikipedia passage. We observe that a private entity occurs 9 times on average in gold training data passages while a public entity appears 4 times on average. There are 22.6k unique private entities in the gold training data passages, and 12.8k unique public entities. \textbf{Passage Length:} Finally, the average length of Wikipedia passages is shorter than the average length of email passages (Table \ref{tab:mutlihopqa_stats}).\footnote{The information density is observationally lower in an email than Wikipedia passage, so we include longer passages to help crowdworkers generate meaningful questions.} Figure \ref{fig:umap} visually shows the distributions of questions and passages.

\begin{figure}
    \centering
    \includegraphics[width=0.65\linewidth]{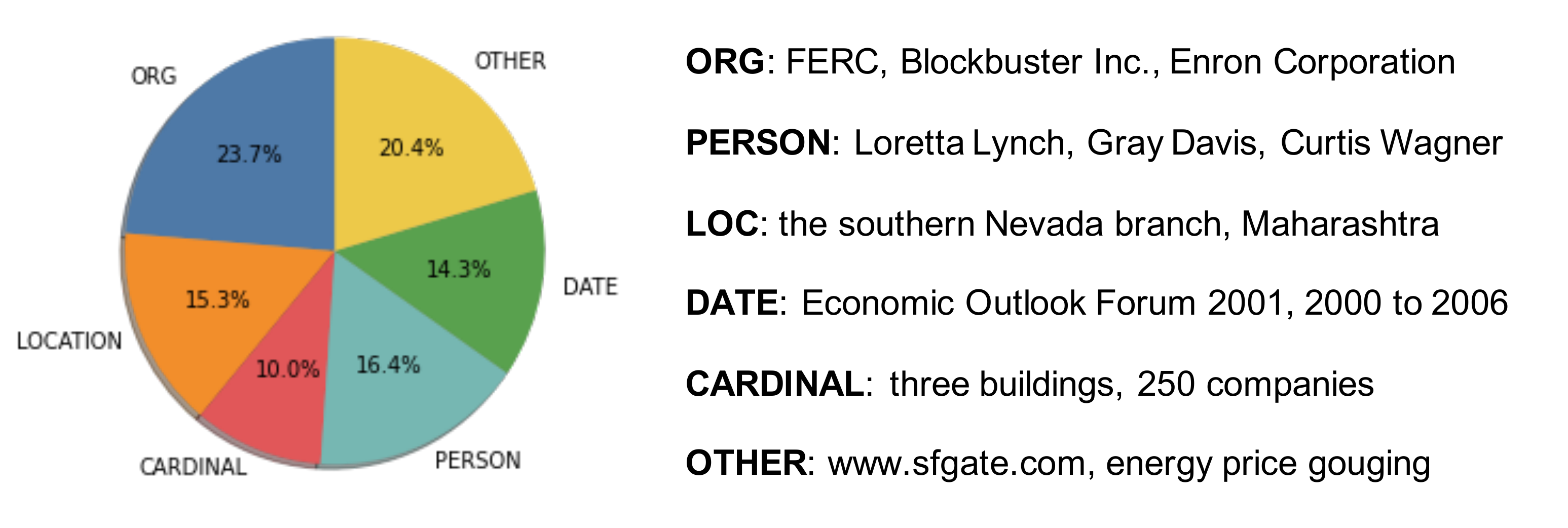}
    \caption[width=0.9\linewidth]{NER-types of answer spans in \datasetname.}
    \label{fig:answer_analysis}
\end{figure}

\paragraph{Answer Types} \datasetname is a factoid QA task so answers tend to be short spans of text containing nouns, or entity names and properties. Figure \ref{fig:answer_analysis} shows the distribution NER tags across answers and examples from each category.

\subsection{Benchmark Limitations}
\label{sec:limits_bench}
\datasetname, like HotpotQA, faces the limitation that crowdworkers see the gold supporting passages when creating questions, which can result in textual overlap between substrings in the questions and passages \cite{trivedi2020dire}. We mitigate these effects through our validation task, and by limiting the allowable degree of overlap between passage pairs and questions through the frontend interface during the generation stage. Further, our questions are not organic user searches as in \citet{47761}, however search logs do not contain questions over public and private data, and existing dialogue systems have not considered retrieval from a private corpus to our knowledge.

Additionally, Enron was a major public corporation and many entities discussed in Enron emails are public entities, so it is possible that public websites and news articles encountered during retriever and reader model pretraining, impact the distinction between public and private questions. We investigate the impact of dataset leakage further in Section \ref{sec:section6}.

\section{Evaluation in the \problemshortname Setting}
\label{sec:solution}
\begin{table*}[t!]
        \begin{center}
    \normalsize
    \begin{tabular}{lcc|cc}
    \toprule
       \multirow{2}{*}{Model}  &      \multicolumn{2}{p{2.7cm}}{\centering \textsc{HotpotQA-\problemshortname}} &     \multicolumn{2}{p{2.7cm}}{\centering \datasetname}  \\
         &    \emph{EM}  &   \emph{F1}    & \emph{EM} & \emph{F1}\\

    \midrule
     No Privacy Baseline         & 62.3 & 75.3 & 45.0 & 53.0  \\
     No Privacy Multi-Index      & 62.3 & 75.3 & 45.0 & 53.0 \\
     Document Privacy            & 56.3 & 68.3 & 36.1 & 43.0 \\
     Query Privacy               & 33.1 & 42.5 & 19.1 & 23.8\\
    \bottomrule
    \end{tabular}
    \normalsize
    \caption{Multi-hop QA datasets using the dense retrieval baseline (MDR) under each privacy setting.}
    \vspace{-2mm}
    \label{tab:benchmarks}
    \end{center}
\end{table*}

\textit{Research Question} How how do existing multi-hop ODQA systems perform under the \problemshortname framework on the HotpotQA-\problemshortname proxy and \datasetname benchmark described in Section \ref{sec:benchmark}.

\paragraph{Model} We use the multi-hop QA method, multi-hop dense retrieval (MDR) \cite{xiong2021mdr}, as the baseline for evaluation, given its simplicity and competitive performance, however the privacy analysis applies to all iterative multi-hop methods. MDR is a dense-passage-retrieval (DPR) bi-encoder model, consisting of a query encoder $g(\cdot)$ and passage encoder $h(\cdot)$ \cite{karpukhin2020dpr}. The model is trained contrastively on tuples of queries, positive passages (containing the answer to the query), and negative passages. For fast retrieval, document embeddings are typically stored in an index designed for efficient similarity search and embedding clustering \cite{faiss2017}. In the first iteration of MDR, the embedding for query $q$ is used to retrieve the $k$ documents $d_1, ..., d_k$ with the highest \textit{retrieval score} according to a maximum inner product search over the dense corpus: 
{\begin{align}
    P(d_i|q) = \frac{\exp(h(d_i)g(q))}{\sum_{d \in D_G} \exp(h(d)g(q))}
\end{align}}

Retrieved documents are each appended to $q$, and the embedding of the resulting $q | d_i$ is used to collect $k$ passage sets of $k$ passages each for the following iteration. The top-$k$ of the $k^2$ passages are again ranked and the top-$k$ are presented to the reader model, which selects a candidate answer in each passage. The candidate with the highest \textit{reader score} is outputted. The reader is a fine-tuned ELECTRA-Large model \cite{clark2020electra}.

\begin{table*}[t]
\small
\begin{tabular}{p{2.7cm}p{12.7cm}}
\toprule
Category & Sample Questions  \\
\midrule
Queries answered 
under \textbf{No Privacy}, but \textit{not} under Document Privacy & \textit{Q1} In which region is the \blue{site of a meeting} between Dabhol manager Wade Cline and Ministry of Power Secretary A. K. Basu located? 

\textit{Q2} What year was the \blue{company that employed Mr. Janac} as general manager founded?

\textit{Q3} What year was the state-owned regulation \blue{board that was in conflict} with Dabhol Power over the 2,184-megawatt DPC project in formed?\\
\midrule
Queries answered under \textbf{Document Privacy} & 
\textit{Q1}  Who was the \blue{deputy campaign manager} in 1992 for California's senior U.S. Senator?

\textit{Q2} The U.S. Representative from New York who served from 1983 to 2013 requested a summary of what \blue{order concerning a price cap complaint}?

\textit{Q3} \blue{How much of the company} now known as the DirecTV Group does General Motors own?
\\
\midrule
Queries answered under \textbf{Query Privacy} \newline\newline  & 
\textit{Q1} The four individuals fired by DWR who had access privilege to ISO control room could not be reached on Friday by who? 

\textit{Q2}  Which CarrierPoint backer also has a partner on the SupplySolution board of directors?

\textit{Q3} At the end of what year did Enron India's managing director responsible for managing operations for Dabhol Power believe it would go online?

\textit{Q4} Who served as the president for the company for which Jerry Meek served as utility operations manager?

\blue{*All evidence is in private emails and not in  Wikipedia.}\\
\bottomrule
\end{tabular}
\caption{Examples of queries answered under different privacy restrictions. \blue{Blue} indicates private information.}
    \vspace{3mm}
    \label{tab:answerable_questions}
\end{table*}

\paragraph{Privacy-Performance Tradeoff} Next we evaluate MDR within the \problemshortname framework. We use question-answering models trained on HotpotQA (i.e. Wikipedia) data, to evaluate performance both on the in-distribution HotpotQA and mixed-distribution \datasetname evaluation data. The latter setting captures the intuition that public and private data will reflect different distributions, and training data is unlikely to be available for private distributions.

\begin{enumerate}
\item \textit{Single-Index Baseline} Here we combine all the public and private documents in a single corpus, setting aside privacy concerns (Table \ref{tab:benchmarks} --- ``No Privacy Baseline'') This is the current standard.
\item \textit{Multiple-Indices} We create two corpora and retrieve the top $k$ from each in each iteration. Note that retrieving less than $k$ documents per index may result in a performance drop vs. the single-index baseline, if the global top-$k$ are all in one corpus. However, instead retrieving the top-$k$ from each, and retaining the top $k_P$ private and $k_G$ public such that $k_P+k_G = k$, we can fully recover the performance of using a single index (Table \ref{tab:benchmarks} --- ``No Privacy Multi-Index'').  The cost of this decision is it introduces up to $2$x as many queries per iteration, since each query is used to retrieve from both indices.
\item \textit{Document Privacy} To maintain document privacy, we cannot use a private passage $p$ retrieved in a prior retrieval iteration to subsequently retrieve from $D_G$. 
Restricting retrievals using $p$ results in a clear performance drop (see Table \ref{tab:benchmarks} --- ``Document Privacy Baseline''). 

\item \textit{Query Privacy} The natural baseline to enforce query privacy is to only retrieve from $D_P$ on each hop. This results in a significant performance drop (see Table \ref{tab:benchmarks} --- ``Query Privacy Baseline'').
\end{enumerate}

We are excitingly able to answer many complex questions \textit{while maintaining privacy} (see examples in Table \ref{tab:answerable_questions} from \datasetname). However at the same time, in maintaining document privacy, the end-to-end question-answering system achieves 9\% worse performance for HotpotQA and 19\% worse performance for \datasetname compared to the quality of the non-private system, and the degradation is even worse if questions are only posed to the private corpus. The performance degradation is undesirable for a deployed system, so our next focus is to investigate the key research challenges towards realizing privacy-preserving retrieval systems.

\section{Challenges in Enabling Public-Private Retrieval}
\label{sec:section6}

In this section, we investigate challenges towards improving the quality of public-private retrieval systems: 
\begin{enumerate}
    \item \textit{Research Question} Can we predict whether a natural language question is unanswerable due to imposed privacy restrictions? We explore this in Selective Prediction, Section \ref{sec:selpred}.
    \item \textit{Research Question} How do retrieval systems perform when public and private data distributions differ? We explore this in Multi-Distribution Retrieval, Section \ref{sec:multiretrieve}.
\end{enumerate}

\subsection{Selective Prediction}
\label{sec:selpred}
To mitigate the privacy-performance tradeoffs observed in Section \ref{sec:solution}, the first natural objective is to answer as many questions as possible (\textit{high coverage}) under imposed privacy constraints, with as high performance as possible (\textit{low risk}).
\paragraph{Design Primitives}
 Ultimately given a multi-hop query $q$, we need to classify between the cases for the $\mathrm{Hop}_i \rightarrow \mathrm{Hop}_{i+1}$ supporting documents where each $\mathrm{Hop} \in \{Private, Public \}$. A question can be answered with \problemshortname-document-privacy so long as the supports are not $Private \rightarrow Public$. A question can be answered with \problemshortname-query-privacy so long as the supports are $Private \rightarrow Private$. \footnote{If a query requires $Private \rightarrow Public$ supporting paragraphs, we can potentially achieve partial query privacy by decomposing the query into multiple single-hop sub-queries and only sending public sub-queries to the public index. Prior work studies query decomposition \cite{min2019multihopdecomp, perez2020qadecomp, wolfson2020break} and we leave the application of decomposition for privacy to future work.}
 To classify between these cases, the options are to use linguistic features and representations, or to use model outputs. Classifying using linguistic features (e.g., entities mentioned) alone is challenging, due to the diversity of user queries. Consider the following HotpotQA examples:
\begin{itemize}
    \item For some queries, required entities are not mentioned by name in the query. E.g., answering \textit{``What screenwriter with credits for `Evolution' co-wrote a film starring Nicolas Cage and Téa Leoni?''} requires retrieving the document for ``The Family Man'' then for ``David Weissman''. 
    \item For other queries, no named entities are mentioned by name and only descriptions of entities are provided. E.g., \textit{``What company claims to manufacture one out of every three objects that provide a shelf life typically ranging from one to five years?''} 
\end{itemize}

Instead, selective prediction \cite{chow1975selectivepred, elyaniv2010selectivepred, geifman2017selectivepred} is a common and general starting point to predict answerability (e.g., \citet{rodriguez2019quizbowl, kamath2020selectiveqa, lewis2021paq}). 

\begin{figure}
    \centering
    \includegraphics[width=0.45\linewidth]{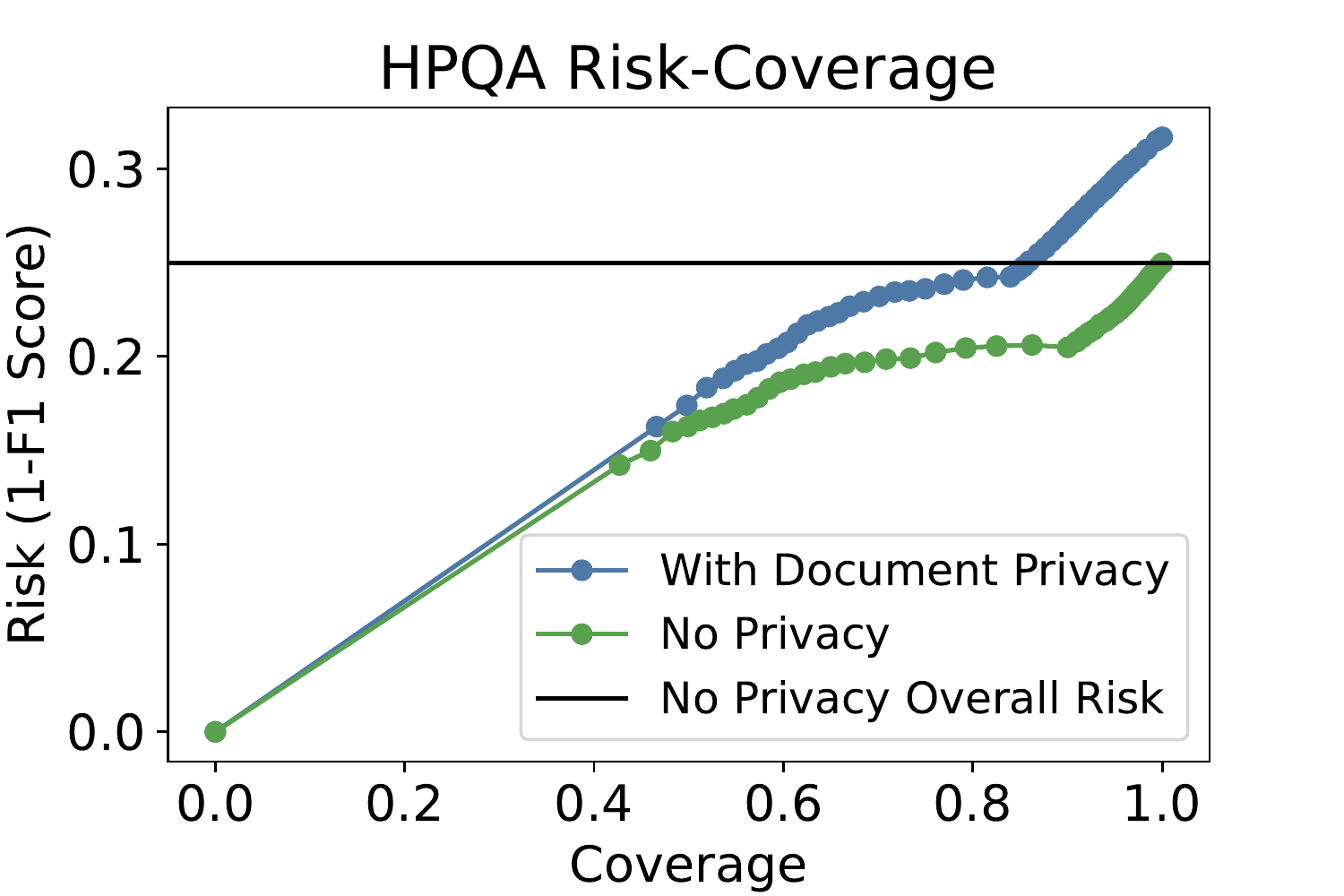}
    \includegraphics[width=0.45\linewidth]{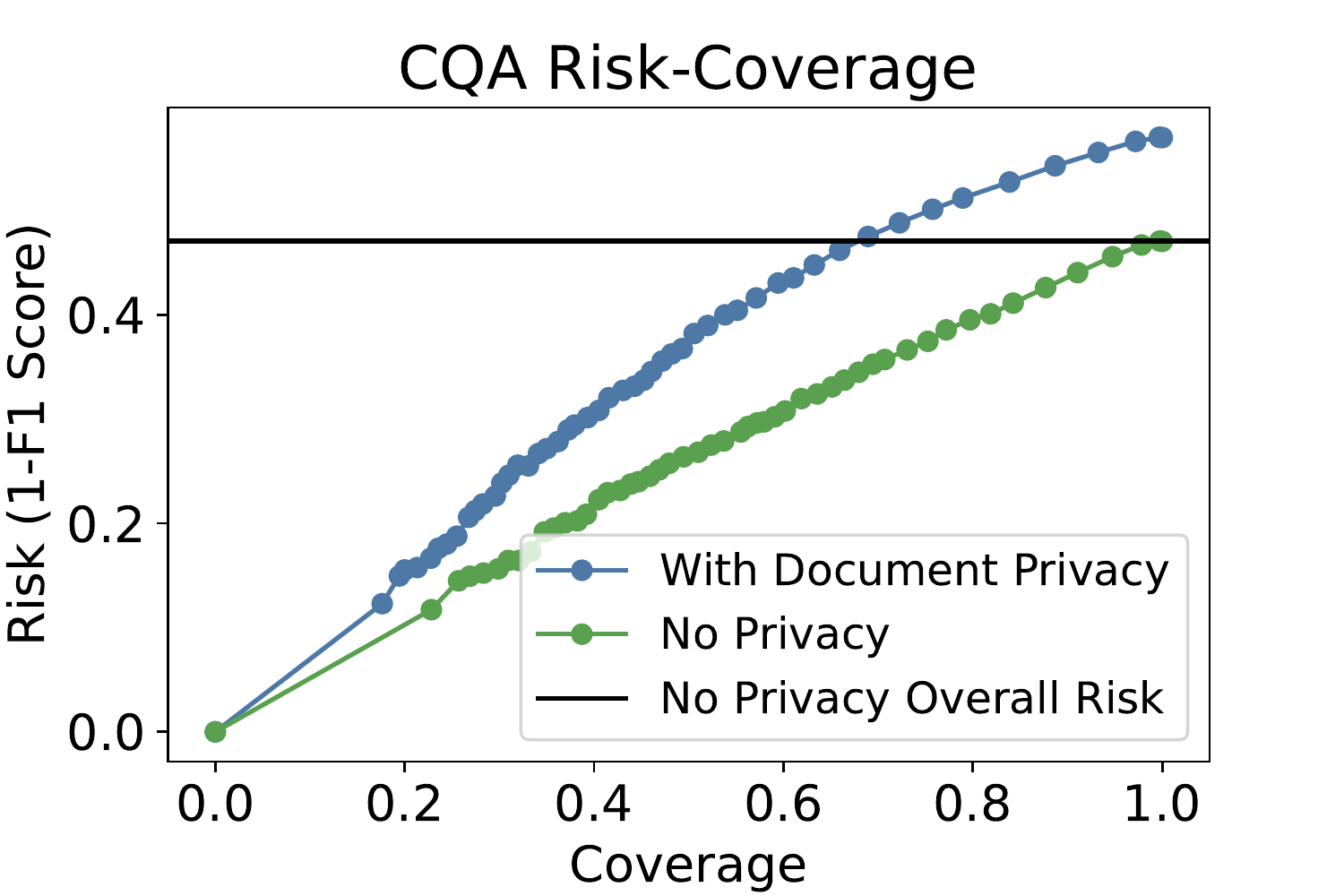}
    \caption[width=0.9\linewidth]{Selective prediction risk-coverage curves using the model trained on HotpotQA data. The left shows results on HotpotQA evaluation data and right on \datasetname test data.} 
    \label{fig:callibration}
\end{figure}

\begin{figure}
    \centering
    \includegraphics[width=0.35\linewidth]{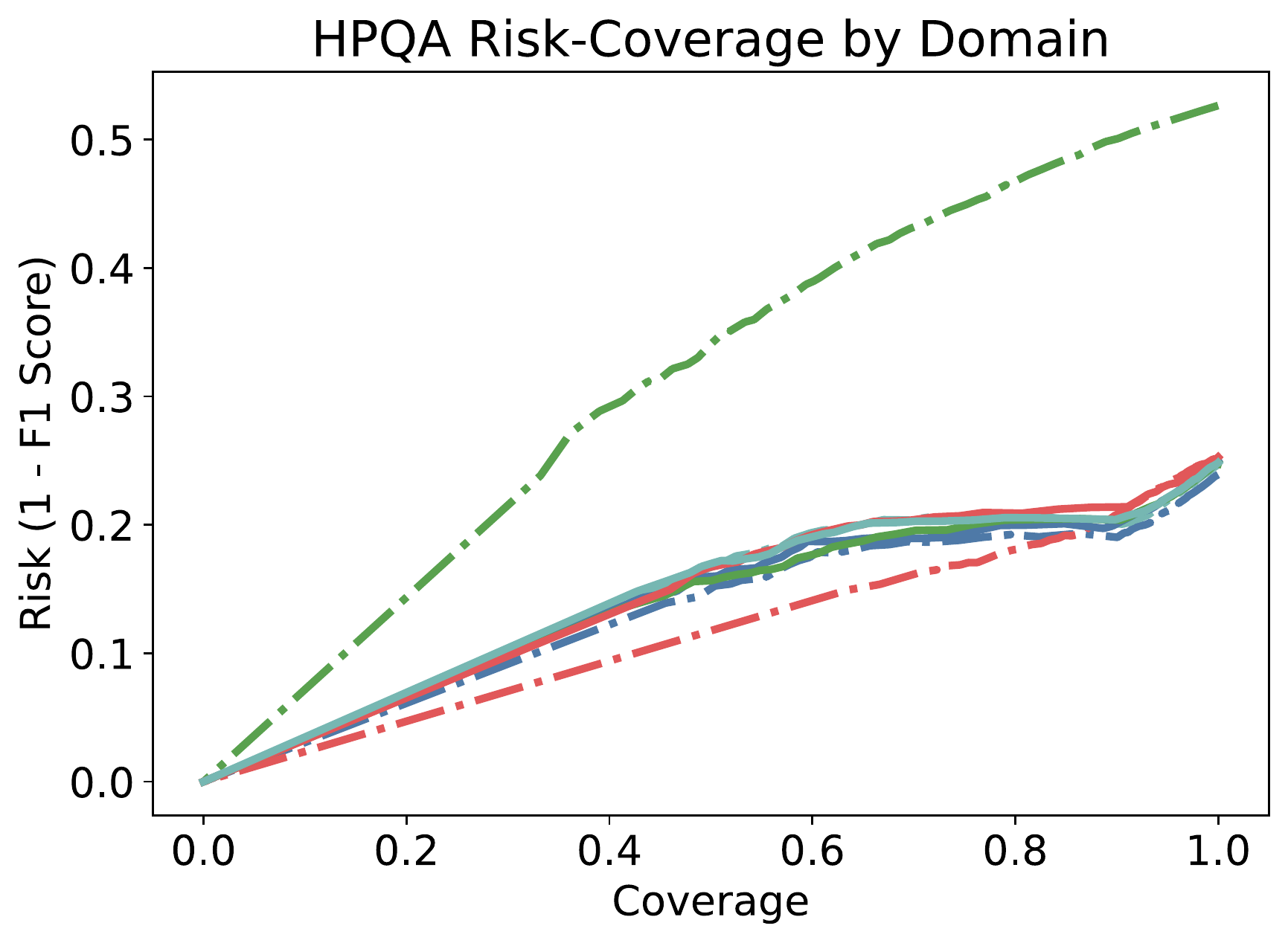}
    \includegraphics[width=0.6\linewidth]{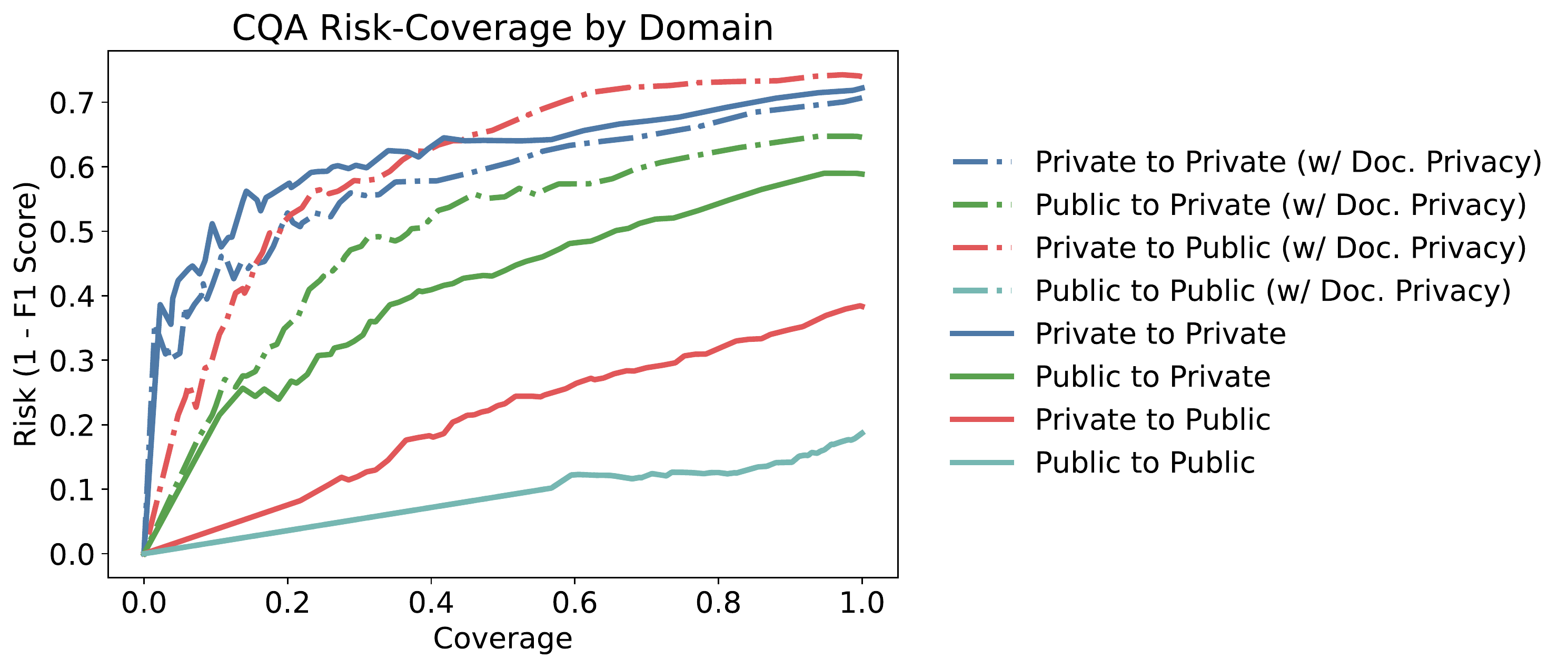}
    \caption[width=0.9\linewidth]{Calibration risk-coverage curves for HotpotQA (left) and \datasetname (right), split by the supporting passage domains for the question at hand. The legend, $\mathrm{Hop_1}$ to $\mathrm{Hop_2}$, indicates the domains of the question's supporting passages. Recall that Document Privacy restricts $Private$ to $Public$ retrieval.} 
    \label{fig:callibration_by_domain}
\end{figure}

In selective prediction, given an input $x$, and a model which outputs $(\hat{y}, c)$, where $\hat{y}$ is the predicted label and $c \in \mathbb{R}$ represents the model's confidence in the prediction, the system provides $\hat{y}$ if $c \geq \gamma$ for some threshold $\gamma \in \mathbb{R}$, and abstains otherwise. 
We evaluate using \textit{risk-coverage} curves \cite{elyaniv2010selectivepred}, where the coverage is the proportion of queries the selective prediction method answers (i.e., examples for which $c \geq \gamma$), and the risk is the error achieved on the covered queries. Intuitively, as $\gamma$ is higher, coverage and risk both tend to decrease. The QA model outputs an answer-string and score for the top $k$ passage chains collected by the retriever, and we compute the softmax over these scores, using the top softmax score as $c$ \cite{hendrycks2017maxprob}.
These are the same reader scores as in Section \ref{sec:solution}; models are trained on HotpotQA data and applied to HotpotQA and \datasetname evaluation data.

\paragraph{Takeaways} 
Figure \ref{fig:callibration} shows the risk-coverage curves for predictions produced after either ``No Privacy'' or ``Document Privacy'' retrieval for HotpotQA (left) and \datasetname (right). The non-private score of 75.3 F1 for HotpotQA is achieved at $85.7$ coverage and 53.0 F1 for \datasetname at $67.8\%$.

\textit{Privacy restrictions appear to increase the selective prediction challenge.} In Figure $3$, the risk-coverage tradeoff is consistently \textit{worse} (i.e., at a given coverage level, the risk is higher) for selective prediction methods applied to the Document Privacy compared to No Privacy baselines. In Figure \ref{fig:callibration_by_domain}, we break down the risk-coverage by the domains of the supporting passages required for $\mathrm{Hop_1} \rightarrow \mathrm{Hop_2}$ on each question. Recall that enforcing Document Privacy restricts $Private \rightarrow Public$ retrieval sequences. We observe the risk-coverage tradeoff worsens not only for $Private \rightarrow Public$, but also for alternate \textit{unrestricted} retrieval paths, such as $Public \rightarrow Private$ for \datasetname (right)) under the Document Privacy vs. No Privacy baseline. Intuitively, if the reader receives low-quality passages for $Private \rightarrow Public$ questions, its confidence may be lower for \textit{similar} $Public \rightarrow Private$ examples. We observe the reader-model's softmax entropy is $38.4\%$ higher across $Public \rightarrow Private$ and $Private \rightarrow Public$ examples in \datasetname when Document Privacy is imposed, compared to the No Privacy baseline. Privacy-restricted examples are essentially out-of-distribution, increasing the selective prediction challenge \cite{kamath2020selectiveqa}.

\textit{Selective prediction quality is much worse for certain sub-distributions of \datasetname.} 
Independent of privacy concerns, Figure \ref{fig:callibration_by_domain} shows worse performance at full-coverage and worse risk-coverage tradeoffs for questions involving private emails. 
Alongside improving predictions of answerability under privacy restrictions, there is significant room to improve
retrieval quality even in the absence of privacy concerns, which we investigate next.

\begin{table*}[t!]
        \begin{center}
    \normalsize
    \begin{tabular}{lcc|cccc}
    \toprule
       \multirow{2}{*}{Retrieval Method}  &      \multicolumn{2}{p{2.7cm}}{\centering \textsc{Overall}} &     \multicolumn{4}{p{4cm}}{\centering Domain-Conditioned}  \\
    & \emph{EM}  &   \emph{F1}    & \emph{EE} & \emph{EW} & \emph{WE} & \emph{WW} \\

    \midrule
    \datasetname-MDR    & 48.9 & 56.5 & 49.5 & 66.4 & 41.8 & 68.3 \\
    HotpotQA-MDR   & 45.0 & 53.0 & 28.7 & 61.7 & 41.1 & 81.3 \\
    Subsampled HotpotQA-MDR  & 37.2 & 43.9 & 23.8 & 51.1 & 28.6 & 72.1 \\
    BM25            & 33.2 & 40.8 & 44.2 & 30.7 & 50.2 & 30.5 \\
    \midrule
    Oracle & 74.1 & 83.4 & 66.5 & 87.5 & 89.4 & 90.4 \\
    \bottomrule
    \end{tabular}
    \normalsize
    \caption{\datasetname results using four retrieval approaches, and oracle retrieval. On the right, we show performance (F1 scores) by the domains of the $\mathrm{Hop_1}$ and $\mathrm{Hop_2}$ gold passages for each question, where email is ``E'' and Wikipedia is ``W''. ``EW'' indicates the $\mathrm{Hop_1}$ gold passage is an email, and $\mathrm{Hop_2}$ gold passage is from Wikipedia.}
    \vspace{-2mm}
    \label{tab:concurrent_baselines}
    \end{center}
\end{table*}

\subsection{Multi-Distribution Retrieval}
\label{sec:multiretrieve}
Progress on the more general multi-domain retrieval problem is an important step towards succeeding on \datasetname and enabling public-private retrieval, as well as retrieval over temporally-evolving data. While in the more common zero-shot retrieval setting \cite{guoa2021multireqa, thakur2021beir} the top $k$ of $k$ passages will be from the out-of-distribution (OOD) corpus for each retrieval, in the underexplored mixed-retrieval setting, it is possible to retrieve zero OOD passages in the top $k$. Each sub-distribution may further benefit from a different retrieval method.

Section 5 shows us that although HotpotQA and \datasetname are curated using the same data collection process, overall performance on \datasetname remains 18.8 F1 worse. Notably, applying models trained on HotpotQA to \datasetname, we observe similar performance on the subset of questions for which the $\mathrm{Hop_1}$ and $\mathrm{Hop_2}$ passages both come from Wikipedia (Table \ref{tab:concurrent_baselines}, 81.3 F1) as the HotpotQA evaluation performance (Table \ref{tab:benchmarks}, 75.0 F1).

\paragraph{Retrieval Baselines} We evaluate on \datasetname using four retrieval baselines: (1) \textbf{\datasetname-MDR} is a dense retriever (MDR, Section 5.1) trained on the \datasetname train set (15.2k examples) and we use this to understand the value of in-domain training data for the task. (2) \textbf{HotpotQA-MDR} is a dense retriever trained on the full HotpotQA train set (90.4K examples) and we use this to understand how well a publicly trained model performs on the public-private mixed distribution. (3) \textbf{Subsampled HotpotQA-MDR} is a dense retriever trained on subsampled HotpotQA data of the same size as the \datasetname train set and we use this to investigate the effect of dataset size. (4) Finally we consider \textbf{BM25} sparse retrieval as prior work indicates its strength in OOD retrieval \cite{thakur2021beir}. Results are in Table \ref{tab:concurrent_baselines}. For each method, $k=100$, where $k$ the number of retrieved passages per hop. We include ablations for different values of $k$, as well as details about the models and experiments, in Appendix \ref{sec:appendix}. The reader for all runs is an ELECTRA-Large model trained on the full HotpotQA training set.

\begin{figure}[t!]
    \centering
    \includegraphics[width=0.5\linewidth]{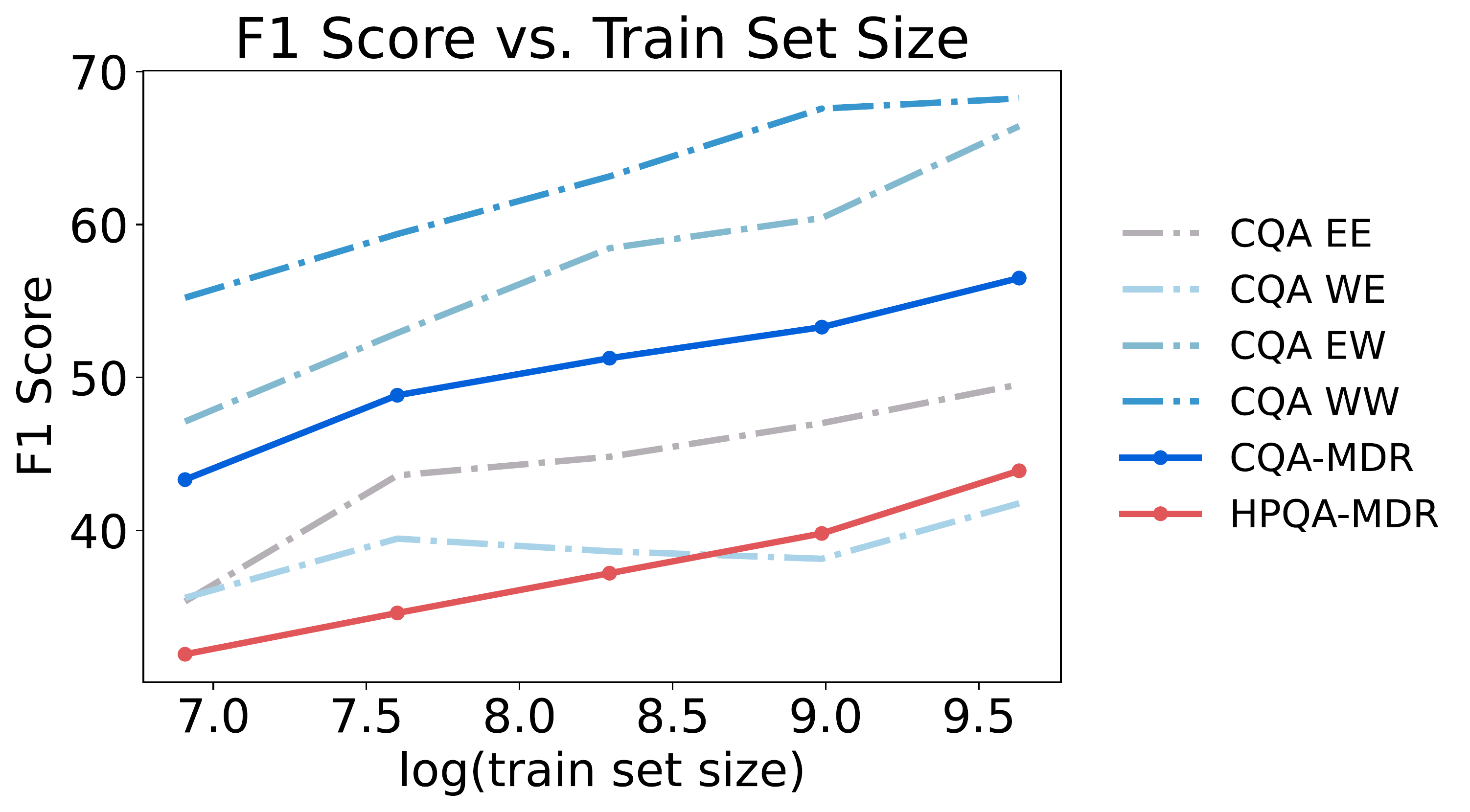}
    \caption[width=\linewidth]{F1 score vs. training dataset size (log scale), training MDR on subsampled HotpotQA (HPQA) and subsampled \datasetname (CQA) training data. We also show trends by the domain of the question's gold passages for CQA.}
    \label{fig:train_size}
    \vspace{-2mm}
\end{figure}

\paragraph{Training Data Size} \textit{Strong dense retrieval performance requires a large amount of training data.} 
Comparing the \datasetname-MDR and Subsampled Hotpot-QA MDR baselines, the former outperforms by 12.6 F1 points as it is evaluated in-domain. However, the HotpotQA-MDR baseline, trained on the full 90k HotpotQA training examples, performs nearly equal to \datasetname-MDR. 
Figure \ref{fig:train_size} shows the performance of Subsampled \datasetname-MDR and Subsampled HotpotQA-MDR for subsample sizes $\in \{1$k, $2$k, $4$k, $8$k$, |\datasetname_\text{Train}|\}$ and the full HotpotQA training dataset. 
Next we observe that the sparse method matches zero-shot performance of using the Subsampled HotpotQA model on \datasetname, but for larger dataset sizes (HotpotQA-MDR) and in-domain training data (\datasetname-MDR), dense retrieval outperforms sparse retrieval.
Notably, it may be difficult to obtain training data for all incurred distributions, especially for private or temporally arising distributions \cite{hawking2004enterprisesearch, chitrita2005personalmetadata}. 

\paragraph{Domain Specific Performance} \textit{Each retrieval method excels in a different subdomain of the benchmark}.  Table \ref{tab:concurrent_baselines} shows the retrieval performance of each method based on whether the gold supporting passages for the first and second hop of the multi-hop example are email (E) or Wikipedia (W) passages. The notation EW means the first hop is an email and second is a Wikipedia passage. HotpotQA-MDR performance on WW questions far exceeds performance on questions where at least one supporting passage is an email. Notably, HotpotQA-MDR gives 81.3 EM for WW, but only 28.7 EM for EE. We also observe that the sparse retriever performs worse than the dense models on Wikipedia-based questions, but better on questions involving an email as $\mathrm{Hop_2}$.
When training on \datasetname, the performance on questions involving emails improves significantly, however remains lower than its performance on Wikipedia-based questions. The WW performance also decreases significantly using \datasetname-MDR. 
We discuss this further in Section 6.3.

\paragraph{Oracle} QA performance using oracle retrieval, i.e., explicitly providing the gold supporting passages to the model, is also provided in Table \ref{tab:concurrent_baselines}. These results demonstrate significant room to improve retrieval, however performance on EE questions also remains low, indicating room to improve the reader as well. 

\paragraph{Dataset Leakage} 
We use RoBERTA-Base for the retriever \cite{liu2019roberta}
and ELECTRA-Large for the reader
\cite{clark2020electra}. As a simple test to investigate the effect of dataset leakage, due to the pretrained language models viewing email data during pretraining, we consider performance using only the Wikipedia passages. The test score is 27.6 EM, where performance is 72.0 EM on WW and 3.3 EM on EE questions. Overall, it is possible that the models may have picked up general knowledge during pretraining that helps reason about Enron concepts. However, these results suggest that access to the private corpus remains important.

\subsection{Error Analysis of Retrieval Methods} 
We conclude with a qualitative discussion of representative errors observed for each retrieval method. 
\paragraph{Dense Retrievers} The primary failure modes we observe for HotpotQA-MDR are: (1) ignoring parts of the question to pick passages reflecting a subset of mentioned entities and details, 
(2) ignoring a short relevant substring within a long $\mathrm{Hop_1}$ email and thus not retrieving the $\mathrm{Hop_2}$ passage successfully,
and (3) choosing Wikipedia passages over email passages. On the slice of examples where the gold $\mathrm{Hop_1}$ passage is an email, 15\% of the time, no emails appear in the top-$k$ $\mathrm{Hop_1}$ results; meanwhile, this only occurs 4\% of the time for $\mathrm{Hop_1}$ Wikipedia. On the slice of EE examples, 64\% of $\mathrm{Hop_2}$ passages are E, while on the slice of WW examples, 99.9\% of $\mathrm{Hop_2}$ passages are W. If we simply \textit{force} equal retrieval from each domain on each hop, we observe up to $2.3$ F1 points improvement on overall \datasetname performance. However, this is a heuristic choice and should be explored further in future work.

Performance on WE questions is notably worse than EW questions and we hypothesize that two factors impact these results: (1) Wikipedia passages generally follow consistent structures,
so it may be easier to retrieve Wikipedia passages on $\mathrm{Hop_2}$ after retrieving Wikipedia on $\mathrm{Hop_1}$, and (2) several emails discuss each Wikipedia-entity, which may increase the noise in $\mathrm{Hop_2}$ (i.e., WE is a one-to-many hop, while for EW, W typically contains one valid entity-specific passage). The latter is intuitively because individuals owning private data truly care about a narrow set of public entities.

\paragraph{Sparse Retrievers} First, we observe the sparse model often ``cheats'' by retrieving the $\mathrm{Hop_2}$ passage, without the $\mathrm{Hop_1}$ passage. For questions where BM25 retrieves the gold $\mathrm{Hop_2}$ passage in the first hop, the score is 64.2 F1, and when this is not the case, the score is 18.3 F1. 

Next, we observe BM25 performance is high on email based questions --- we compute WW questions have an average length of 97 characters, while EE questions have an average length of 141 characters. Perhaps, due to the nature of how the dataset is constructed, namely crowdworkers can see the passages before they write the questions, we may be underestimating the need for skills dense models provide (e.g., fuzzy semantic matching)  and overestimating the quality of sparse models that benefit from direct matching. We observe several other benchmarks reported in \cite{thakur2021beir} on which BM25 outperforms dense retrieval, use similar annotation pipelines during question generation (e.g., \citet{wadden2020factorfiction, yang2018hotpotqa}).



\section{Discussion and Future Work}
\paragraph{Privacy-Preserving Personalized Retrieval Systems} We hope this work inspires interest in realizing the potential of privacy-preserving personalized open-domain systems. Potential future directions include decomposing multi-hop queries into public and private sub-queries to address query-privacy \cite{min2019multihopdecomp, perez2020qadecomp}. Additionally, reformulating queries by including personal keywords could help produce more meaningful retrievals \cite{carpineto2012aqe}. For example, if an ML practitioner asks a question about ``Michael Jordan'' the ML professor, a naive public search may return many passages about the basketball player.\footnote{Google returns 1.4Bn results for ``Michael Jordan'', 170M search results for ``Michael Jordan basketball'', and 85M for ``Michael Jordan AI'', though a large portion of the latter are results about ``Air Jordans''.} Perhaps a reformulated query with keywords such as ``ML'' would yield more relevant results. Future work could study the tradeoff between providing additional personal context in the query versus the number of public passages one would need to retrieve. Overall retrieval is an exciting direction for incorporating personal context, without requiring any training.

\paragraph{Retriever Generalization} While prior work considers zero-shot generalization \cite{thakur2021beir, guoa2021multireqa}, retrieval-based systems over personal or temporally-changing data will need to retrieve from a mixture of in and out-of-distribution  data. In the former setting, $k$ of $k$ retrieved passages will be OOD passages, while in the latter setting, it is possible that few (or $0$) OOD passages are retrieved, for example if the retriever scores are distributionally higher for ID passages. It is also possible that domain labels do not exist for certain retrieval applications. In contrast to using a single retriever for in and OOD data, a system that \textit{routes} questions to different retrievers, depending on question attributes, is another possibility. We hope \datasetname facilitates further study of concurrent multi-domain retrieval.

\section{Conclusion}
This work asks how to personalize retrieval-based systems in a privacy-preserving way and identifies that arbitrary autoregressive retrieval over public and private data poses a privacy concern. In summary, we define the \problemshortname privacy framework, present a new multi-domain multi-hop benchmark called \textit{\datasetname} for the novel retrieval setting, and demonstrate the privacy-performance tradeoffs faced by existing open-domain systems. We finally investigate two challenges towards realizing the potential of public-private retrieval systems: using selective prediction to manage the privacy-performance tradeoff and concurrently retrieving over multiple distributions. We hope this work inspires new privacy-preserving solutions for personalized retrieval-based systems.

\section*{Acknowledgements}
We thank Jack Urbaneck and Wenhan Xiong for answering our questions regarding the Mephisto framework and Multi-hop Dense Retrieval respectively. We thank members of the Hazy Research Lab, Facebook AI Research, and Stanford AI Lab for their helpful feedback and discussions. We gratefully acknowledge the support of NIH under No. U54EB020405 (Mobilize), NSF under Nos. CCF1763315 (Beyond Sparsity), CCF1563078 (Volume to Velocity), and 1937301 (RTML); ARL under No. W911NF-21-2-0251 (Interactive Human-AI Teaming); ONR under No. N000141712266 (Unifying Weak Supervision); ONR N00014-20-1-2480: Understanding and Applying Non-Euclidean Geometry in Machine Learning; N000142012275 (NEPTUNE); NXP, Xilinx, LETI-CEA, Intel, IBM, Microsoft, NEC, Toshiba, TSMC, ARM, Hitachi, BASF, Accenture, Ericsson, Qualcomm, Analog Devices, Google Cloud, Salesforce, Total, the HAI-GCP Cloud Credits for Research program,  the Stanford Data Science Initiative (SDSI), Stanford Graduate Fellowship, and members of the Stanford DAWN project: Facebook, Google, and VMWare. The U.S. Government is authorized to reproduce and distribute reprints for Governmental purposes notwithstanding any copyright notation thereon. Any opinions, findings, and conclusions or recommendations expressed in this material are those of the authors and do not necessarily reflect the views, policies, or endorsements, either expressed or implied, of NIH, ONR, or the U.S. Government.

\bibliographystyle{unsrtnat}
\bibliography{references}  

\appendix
\label{sec:appendix}

\section{Model Details}
This section provides details about the retrieval and reader models and experiment settings. Experiments are conducted on 8 NVidia-A100 GPUs.

\subsection{Dense Retrieval} We use the model implementations for the MDR (dense retriever) provided by \citet{xiong2021mdr}.\footnote{\url{https://github.com/facebookresearch/multihop_dense_retrieval}} For the non-private experiments, we use the base retrieval algorithm; we extend the base implementation for the private-retrieval modes described in Section 3 and release our implementation. We construct the dense passage corpus using FAISS \cite{faiss2017}, and use exact inner product search as in the original implementation.

The retriever is trained with a contrastive loss as in \citet{karpukhin2020dpr}, where each query is paired with a (gold annotated) positive passage and $m$ negative passages to approximate the softmax over all passages. We consider two methods of collecting negative passages: first, we use random passages from the corpus that do not contain the answer (random), and second, we use one top-ranking passage from BM25 that does not contain the answer as a hard-negative paired with remaining random negatives. We do not observe a large difference between the two approaches for \datasetname-results (also observed in \cite{xiong2021mdr}), and thus use random negatives for all experiments. We hope to experiment with additional methods of selecting negatives for \datasetname in future work.

The number of retrieved passages per retrieval, $k$, is an important hyperparameter as increasing $k$ tends to increase recall, but sacrifice precision. Using larger values of $k$ is also less efficient at inference time. We use $k=100$ for all experiments in the paper and Table \ref{tab:mdr_varyk} shows the effect of using different values of $k$ on retrieval performance (HotpotQA-MDR, \datasetname eval data).

\begin{table}[t!]
    \begin{center}
    \normalsize
    \begin{tabular}{llcc}
    \toprule
    Model  &    Avg-PR   & F1 \\
    \midrule
    $k=1$       & 41.4  & 33.5 \\
    $k=10$      & 55.9 &  44.7 \\
    $k=25$      & 63.3  & 48.0 \\
    $k=50$      & 68.4  & 50.4 \\
    $k=100$     & 73.8  & 53.0 \\
    \bottomrule
    \end{tabular}
    \normalsize
    \caption{Retrieval performance (Average Passage-Recall@k, F1) for $k \in \{1, 10, 25, 50, 100\}$ retrieved passages per hop using the retriever trained on HotpotQA for OOD \datasetname test data.}
    \vspace{2mm}
    \label{tab:mdr_varyk}
    \end{center}
\end{table}
\paragraph{Inference-Only} For the MDR experiments in Section 6, and the HotpotQA-MDR experiments in Section 5, we use an MDR-model trained in the Wikipedia domain (i.e., HotpotQA training data) to retrieve passages for the the HotpotQA-\problemshortname and \datasetname evaluation sets. For these experiments, we directly use the provided question encoder and passage encoder checkpoints. 

\paragraph{Training and Inference} For the \datasetname-MDR and Subsampled HotpotQA-MDR experiments, we train the MDR model from scratch, finding the hyperparameters in Table \ref{tab:hyperparams_retrieval} work best.
\begin{table}[t!]
    \begin{center}
    \normalsize
    \begin{tabular}{llcc}
    \toprule
    Model  &    Avg-PR    \\
    \midrule
    Learning Rate                       & 5e-5 \\
    Batch Size                          & 150   \\
    Maximum passage length              & 300  \\
    Maximum query length at initial hop & 70   \\
    Maximum query length at 2nd hop     & 350  \\
    Warmup ratio                        & 0.1  \\
    Gradient clipping norm              & 2.0 \\
    Traininig epoch                     & 64   \\
    Weight decay                        & 0    \\
    \bottomrule
    \end{tabular}
    \normalsize
    \caption{Retrieval hyperparameters for MDR training on \datasetname and Subsampled-HotpotQA experiments.}
    \vspace{2mm}
    \label{tab:hyperparams_retrieval}
    \end{center}
\end{table}

\subsection{Sparse Retrieval}
\begin{table}[t!]
    \begin{center}
    \normalsize
    \begin{tabular}{llcc}
    \toprule
    Model  &    F1   \\
    \midrule
    $k=1$      &  22.0  \\
    $k=10$      & 34.6  \\
    $k=25$      & 37.8  \\
    $k=50$      & 39.3  \\
    $k=100$     & 40.8  \\
    \bottomrule
    \end{tabular}
    \normalsize
    \caption{F1 score on the \datasetname test data  for $k \in \{1, 10, 25, 50, 100\}$ retrieved passages per hop using BM25 sparse retrieval.}
    \vspace{2mm}
    \label{tab:bm25_varyk}
    \end{center}
\end{table}
For the sparse retrieval baseline, we use the Pyserini BM25 implementation using default parameters.\footnote{\url{https://github.com/castorini/pyserini}} We consider different values of $k \in \{1, 10, 25, 50, 100\}$ per retrieval and report the retrieval performance in Table \ref{tab:bm25_varyk}. We generate the second hop query by concatenating the text of the initial query and first hop passages.

\subsection{QA Model}
We use the provided ELECTRA-Large reader model checkpoint from \citet{xiong2021mdr} for all experiments. The model was trained on HotpotQA training data. Using the same reader is useful to understand how retrieval quality affects performance, in the absence of reader modifications.

\begin{table}[t!]
    \begin{center}
    \normalsize
    \begin{tabular}{llcc}
    \toprule
    Model  &    EM  \\
    \midrule
    Single-hop $\text{FiD}_{\text{base}}$    & 30.3  \\
    Multi-hop $\text{FiD}_{\text{base}}$     & 55.9 \\
    Single-hop $\text{FiD}_{\text{large}}$   & 35.3 \\
    Multi-hop $\text{FiD}_{\text{large}}$ *  & 61.7 \\
    \bottomrule
    \end{tabular}
    \normalsize
    \caption{Here we ask how many hops are required to answer the benchmark multi-hop questions. We the same checkpoint MDR models trained on HotpotQA to retrieve hop 1 and hop 2 passages. We train the reader model on only hop 1 passages (Single-hop) and compare to the performance of using hop 1 and hop 2 passages (Multi-hop). We provide results using FiD with T5-base and T5-large. *Reported in \cite{xiong2021mdr}.}
    \vspace{2mm}
    \label{tab:fid}
    \end{center}
\end{table}

\subsection{Are two hops necessary?} The document privacy challenge is a consequence of the autoregressive retrieval process. Here we ask whether two hops are in fact necessary to answer multi-hop benchmark questions. 

In the \textit{Single-hop} baseline, we use the MDR model to retrieve $k=50$ passages, but stop after the first hop. We train and evaluate a Fusion-in-Decoder (FiD) model \cite{izacard2021fid} on the resulting contexts. To motivate the choice of reader, FiD \textit{combines} information across multiple passages simultaneously, whereas ELECTRA searches for answer spans individually in each passage. We compare performance to the \textit{Multi-hop baseline}, where we take the top $50$ passage chains from the same MDR model, concatenate the two passages in each chain to obtain 50 contexts, and train and evaluate a Fusion-in-Decoder model on the resulting data (as in \citet{xiong2021mdr}). We observe that the multi-hop baseline performs 26.4 F1 points higher, indicating the benefit of multiple iterations. See results in Table \ref{tab:fid}.

We train the FiD models for 15000 steps, with a learning rate of $5e-05$, per-GPU batch size of $1$, maximum text length of $250$ when using one passage and $512$ for two passages, and maximum answer length of $20$, for one random seed.

\begin{table*}[t!]
    \begin{center}
    \normalsize
    \begin{tabular}{llccc}
    \toprule
    Benchmark   &  Model  &    EM  &   F1  \\
    \midrule
    \multirow{4}{*}{\datasetname}
    &  No Privacy Baseline         & 49.3 & 55.8  \\
    &  Multi-Index Baseline    & 49.3 & 55.8  \\
    &  Document Privacy Baseline & 38.6 & 45.0   \\
    &  Query Privacy  Baseline  & 19.1 & 23.9   \\
    \bottomrule
    \end{tabular}
    \normalsize
    \caption{Multi-hop QA datasets using MDR under each privacy setting. Here we include results for the \datasetname Dev split.}
    \vspace{-2mm}
    \label{tab:benchmarks_dev}
    \end{center}
\end{table*}

\section{Additional Details for \problemshortname Baselines}

\begin{figure}[t!]
    \centering
    \includegraphics[width=0.33\linewidth]{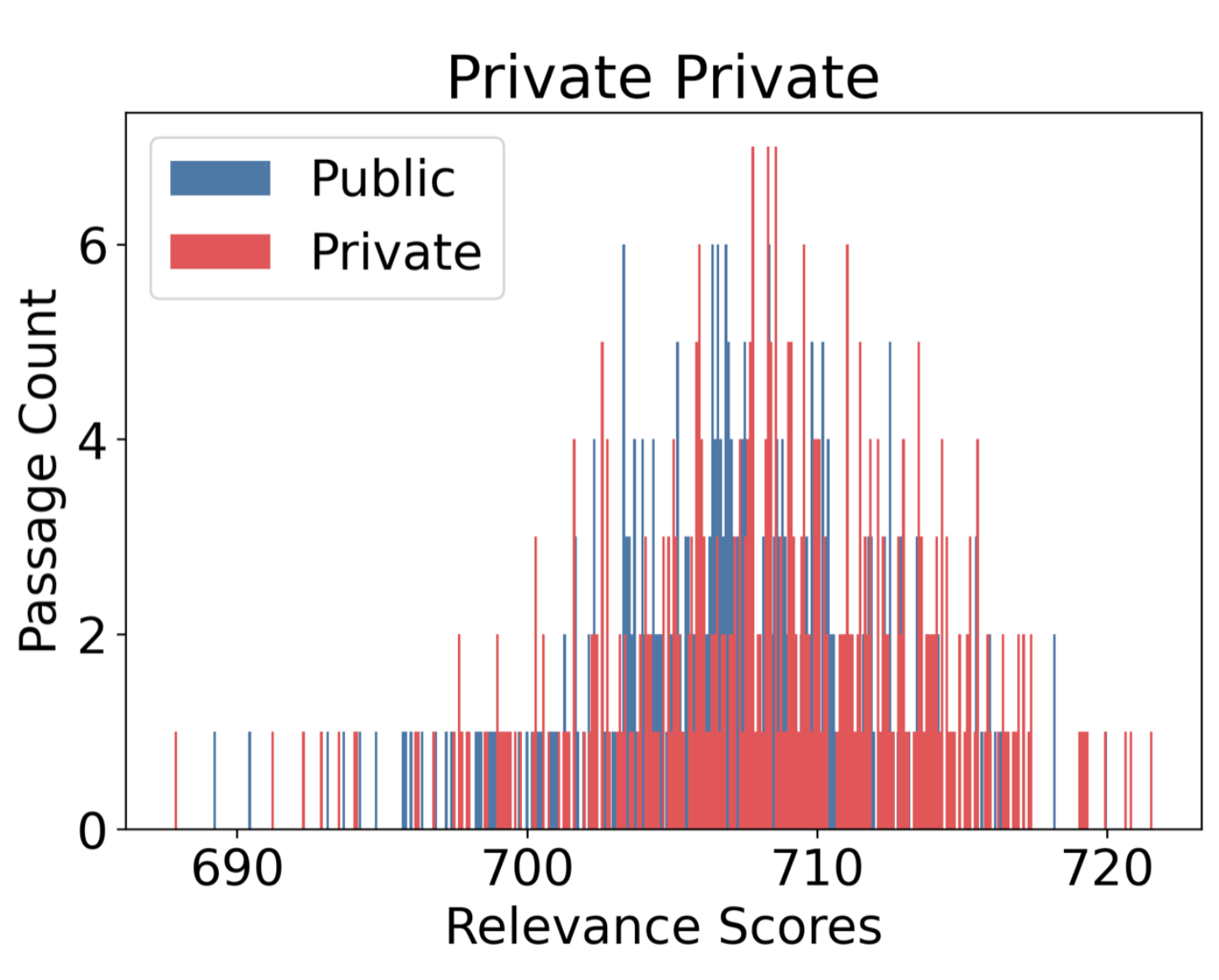}
    \includegraphics[width=0.33\linewidth]{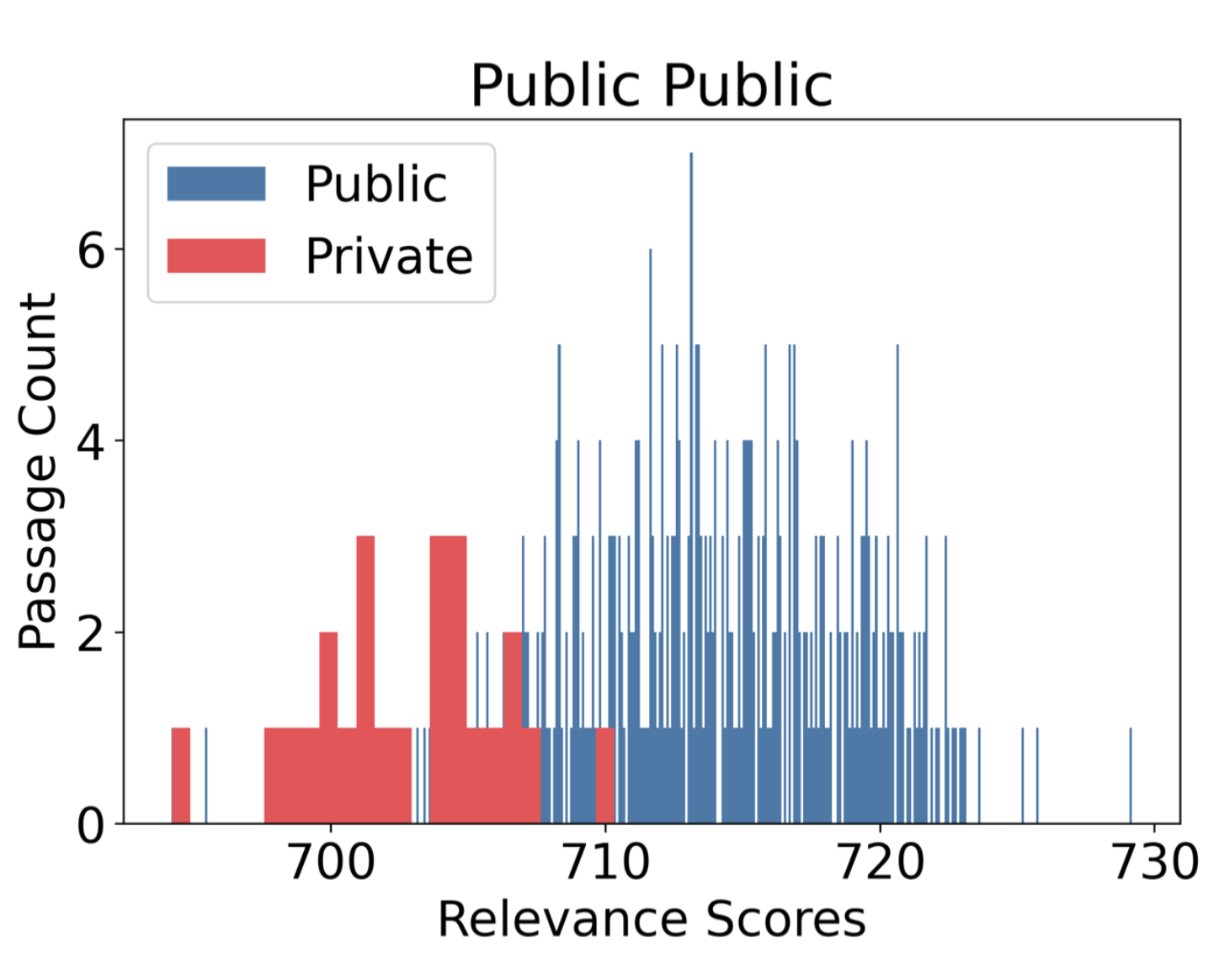}
    \caption[width=0.9\linewidth]{Number of passages retrieved in $\mathrm{Hop_1}$ by relevance score, for each type of \datasetname question, based on the gold supporting passage types.} 
    \label{fig:relevancescores}
\end{figure}

In Table \ref{tab:benchmarks_dev}, we provide QA results for the \datasetname Dev split. 

Figure \ref{fig:relevancescores} show that there is a clear separation between the relevance score distributions from the email vs. Wikipedia corpus for questions based on Wikipedia (public) passages, but this is is not the case for questions based on email passages. The relevance score distributions are not-necessarily well-aligned in the mixed-distribution retrieval setting, contributing to the difficulty and difference vs. zero-shot retrieval.

\section{Additional \datasetname Analysis}

\begin{figure*}[t!]
    \centering
    \includegraphics[width=0.9\linewidth]{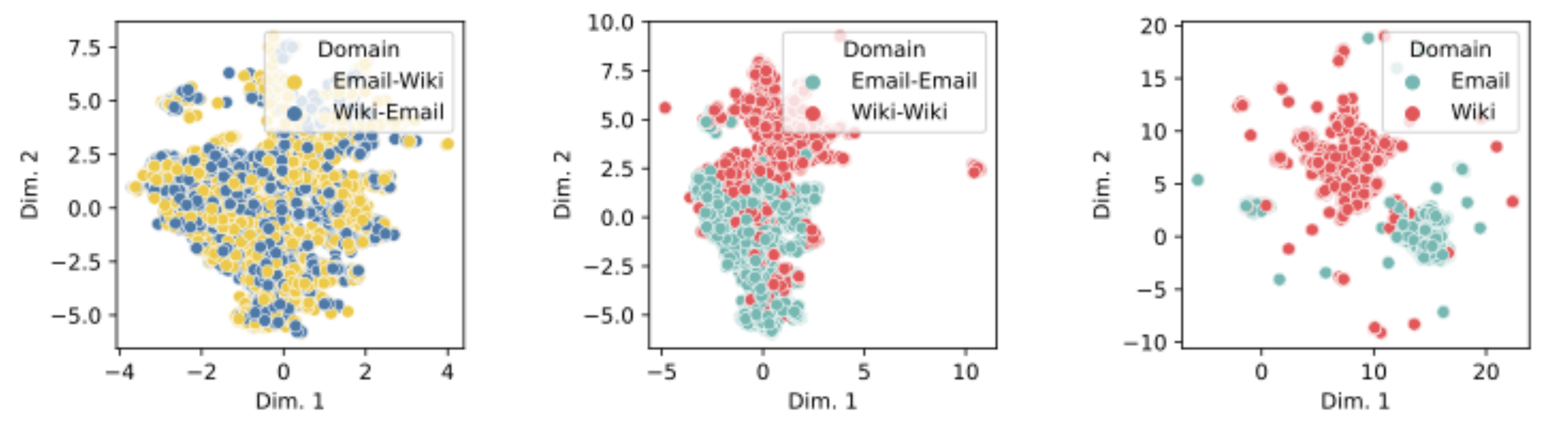}
    \caption[width=0.85\linewidth]{UMAP of BERT-base  
    embeddings, using \citet{reimers-2019-sentence-bert}, of \datasetname \textbf{questions} based on the domains of the gold passage chain to answer the question (left and middle). I.e., questions that require an Email passage for hop 1 and Wikipedia passage for hop 2 are shown as ``Wiki-Email''. Embeddings for all gold \textbf{passages} are also shown, split by domain (right).}
    \label{fig:umap}
\end{figure*}

Figure \ref{fig:umap} (Left, Middle) shows the UMAP plots of \datasetname questions using BERT-base representations, split by whether the gold hop passages are both from the same domain (e.g., two Wikipedia or two email passages) or require one passage from each domain. The plots reflect a separation between Wiki-based and email-based questions and passages.

\begin{table}[t]
\begin{center}
\begin{tabular}{p{5cm}p{1.5cm}}
\toprule
Avg. words per question & 28 \\
Avg. words per Email passage & 149 \\
Avg. words per Wiki passage   & 44 \\
Avg. words per answer & 2 \\
\bottomrule
\end{tabular}
\caption{Length statistics for \datasetname.}
\vspace{3mm}
\label{tab:mutlihopqa_stats}
\end{center}
\end{table}

\begin{table}[t]
\begin{center}
\begin{tabular}{p{1cm}p{1cm}p{0.75cm}p{0.75cm}p{0.75cm}p{0.75cm}}
\toprule
Split & Total & EE & EW & WE & WW \\
\midrule
Train & 15,239  &  3762 & 4002 & 3431 & 4044 \\
Dev   & 1,600 & 400 & 400 & 400 & 400 \\
Test  & 1,600 & 400 & 400 & 400 & 400 \\
\bottomrule
\end{tabular}
\caption{\datasetname Distribution over the gold-passage domains required for benchmark questions. Domains are emails (E) and Wikipedia (W), and ``EW'' indicates the hop 1 gold passage is from Enron and hop 2 gold passage is from Wikipedia. The evaluation sets are balanced between questions with gold passages as emails vs. Wikipedia passages for hops 1 and 2.}
\vspace{3mm}
\label{tab:mutlihopqa_domaintypes}
\end{center}
\end{table}

\begin{table*}[t]
\begin{tabular}{p{15cm}}
\toprule
\blue{Example 1: shows how a question can be answered by an alternate retrieval path than the gold path. The \textit{Alternate Hop 1} passage also depicts typos which are more prevalent in Enron compared to Wikipedia passages.}
\newline

\textit{Multi-hop Question} Reliant Energy is based in a city located in which Texas county? \newline

\textit{Gold Hop 1 (Email)} \textbf{Reliant Energy of Houston}, another company that resisted demands for business records, on Wednesday signed a confidentiality agreement with Dunn's committee and will begin bringing 250,000 documents to a depository in Sacramento, said Reliant spokesman Marty Wilson. Dunn said other companies have begun to deliver documents to Sacramento, but not all are fully complying with subpoenas. ...\newline

\textit{Alternate Hop 1 (Email)} ... "Independent power generators have come under increasing scrutiny and are being investigated by the state's Attorney General Bill Lockyer's office fo r gaming the market." Generators are being investigated as to whether they hav e shut plants for maintenance in order to spike prices during peak periods an d periods when the California Independent System Operator declares alerts whe n power reserves drop below certain levels in the state. \textbf{Reliant Energy is based in Houston, Texas}....
\\ 
\midrule
\blue{Example 2: shows how the same email passage can cover multiple topics.} In contrast to Wikipedia, where passages are about a single entity, other types of documents including emails can cover many topics in the same passage. Thus, the single dense embedding generated per passage in retrieval methods such as DPR may not be as effective. This is a \textit{bridge} question. \newline
 
\textit{Multi-hop Question} How much power can the company reported on October 1 2001 to be in talks to acquire an Indian Enron stake generate?\newline

\textit{Gold Hop 1 (Email)} World Watch The Wall Street Journal, 10 01 01 INDIA: Panel suggests Indian govt pay in Enron row-paper. Reuters English News Service, 10 01 01 INDIA: \textbf{Tata Power said in talks to buy India Enron stake}. Reuters English News Service, 10 01 01 Greece Awards 4 Electricity Production, 8 Supply Permits ... Portland Oregonian, 09 29 01 Firms Push Edison Near Bankruptcy Energy...\newline

\textit{Gold Hop 2 (Wiki)} The Tata Power Company Limited is an Indian electric utility company based in Mumbai, Maharashtra, India and is part of the Tata Group. The core business of the company is to generate, transmit and distribute electricity. With an installed \textbf{electricity generation capacity of 10,577MW}, it is India's largest integrated power company. At the end of August 2013, its market capitalisation was \$2.74 billion.\\
\midrule
\blue{Example 3: shows an example requiring \textit{list-based reasoning}. This occurs in several benchmark questions. This is a \textit{bridge} question.} \newline

\textit{Multi-hop Question} The seven economic commentators at Economic Outlook Forum 2001 were Ben Hermalin's co-chair, Severin Borenstein, Jerry Engel, Rich Lyons, Ken Rosen, Janet Yellen, and a professor who was born in what year?\newline

\textit{Gold Hop 1 (Email)} Dear Haas Evening MBA Students, On Friday afternoon November 9, 2001, some of the School's most distinguished economists and I will participate in a "teach-in" about the US economy. "Economic Outlook Forum 2001" will examine ... I am fortunate to co-chair this session with Professor Ben Hermalin, who will begin serving as Interim Dean of the Haas School in January 2002. Professor Hermalin will moderate the panel presentations and following discussion. In addition to myself, our \textbf{economic commentators will be Professors Severin Borenstein, Jerry Engel, Rich Lyons, Ken Rosen, Hal Varian, and Janet Yellen}...\newline

\textit{Gold Hop 2 (Wiki)} Hal Ronald Varian (\textbf{born March 18, 1947} in Wooster, Ohio) is an economist specializing in microeconomics and information economics. He is the chief economist at Google and he holds the title of emeritus professor at the University of California, Berkeley where he was founding dean of the School of Information. He has written ...
\\
\bottomrule
\end{tabular}
\caption{Illustrative examples of properties of \datasetname.}
    \label{tab:additional_examples}
\end{table*}

\begin{table*}[t]
\begin{tabular}{p{15cm}}
\toprule
\blue{Example 4: shows an example of an \textit{attribute} style question, in which both passages provide an attribute about the same entity (i.e., ``Idealab!'').}\newline

\textit{Multi-hop Question} Funding Metiom filed for Chapter 11 after investors backed out of which company founded by Bill Gross in 1996?\newline

\textit{Gold Hop 2 (Email)} ... DigiPlex Raises \$48 Million Equity, \$35 Million Debt STSN Gets \$66.5M of Series D Debt and Equity Tribune Media Services Takes Majority Stake in TVData Viator Closing \$5M to \$10M Series C Round in Next Two Weeks bad news WorkingWoman.com Lays Off 63\%; Looking for Buyers, Funding \textbf{Metiom Files for Chapter 11 after Investors Back Out Idealab!}\newline

\textit{Gold Hop 2 (Wiki)} \textbf{Idealab was founded by Bill Gross (not the same Bill Gross of PIMCO) in March 1996.} Prior to Idealab, he founded GNP Loudspeakers (now GNP Audio Video), an audio equipment manufacturer; GNP Development Inc., acquired by Lotus Software; and Knowledge Adventure, an educational software company, later acquired by Cendant... \\
\midrule
\blue{Example 5: shows an example of a \textit{yes-no} style question. For these questions (a subset of the comparison questions), the answer is not a span in the passages.}\newline

\textit{Multi-passage Question} Did the company who appointed Carol S. Schmitt as vice president secure all of its expected first round of funding?\newline

\textit{Passage 1} ... Fabless Semiconductor Firm Secures \$8.2 Million in Round One AGOURA HILLS, Calif. -- Internet Machines, a fabless semiconductor company that develops software and services for data communications markets, said it secured \$8.2 million in its first round of funding. ...  Management App Firm Gets \$5 Million of \$8 Million Round One CAMBRIDGE, Mass. -- \textbf{Bluesocket, which develops management software for Bluetooth-enabled networks, said it secured \$5 million of its expected \$8 million first round of funding} from St. Paul Venture Capital and Osborn Capital.\newline

\textit{Passage 2} ... \textbf{Bluesocket, which develops security and management products for wireless local area networks, said it appointed Carol S. Schmitt as vice president} of business development. Prior to joining the company, Ms. Schmitt was a business and market development consultant in Los Gatos, Calif. Bluesocket is backed by Osborn Capital and St. Paul Venture Capital.
\\
\midrule
\blue{Example 6: shows an example of a non yes-no \textit{comparison} style question. For these questions, the answer is the one of the two entities being compared, where one entity appears in each passage.}\newline

\textit{Multi-passage Question} Which company out of Regency Capital and StellaService started its business operations first? \newline

\textit{Passage 1} ... NEW YORK (VENTUREWIRE) -- Privacy Protection, which does business as Eprivex.com and is a developer of electronic privacy technology and persona l privacy protection services, said it must cease operations unless it can complete its seed round of \$1.5 million, wholly or incrementally, from individual or private investors. \textbf{The company, which was founded in March 2000, has received prior financing from individual investors including Roge r Dietch, founder of Regency Capital,} as well as from Jesse L. Martin, Jerry Orbach, and Sam Waterston, all of whom are actors on the NBC television sho w Law and Order. \newline

\textit{Passage 2} StellaService Inc. is a privately held American information and measurement company with headquarters in New York City (USA). The company measures and rates the customer service performance of online companies in a process audited by global accounting and auditing firm KPMG. \textbf{Founded in 2009,} it produces both Stella Metrics (a mystery shopping platform) and Stella Connect (a customer feedback system).
\\
\bottomrule
\end{tabular}
\vspace{3mm}
\end{table*}

In Table \ref{tab:mutlihopqa_domaintypes}, we provide statistics for the number of \datasetname questions that require gold supporting passages from each set of privacy scopes.

In Table \ref{tab:additional_examples} we provide additional examples to illustrate key properties and question types (i.e., bridge, attribute, and comparison) appearing in \datasetname.

\section{Additional Details on the Creation of \datasetname }
\subsection{Data Preprocessing}
\paragraph{Public Wikipedia Data Preprocessing} We use the same corpus of $5.2$ million Wikipedia passages from \citet{xiong2021mdr} as public data.\footnote{\url{https://github.com/facebookresearch/multihop\_dense\_retrieval}}
We use NLTK \cite{bird2009nltk} for sentence tokenization --- this is important for storing the indices of supporting sentences.

\paragraph{Private Enron Data Preprocessing}
We download the May 7, 2015 version of the Enron Emails dataset distributed by CMU. \footnote{\url{www.cs.cmu.edu/~enron/}} We select the ``Jeff Dasovich'' inbox as the personal data source because the inbox is amongst the largest ($28,234$ emails) and the employee was a ``Government Relation Executive'', so the emails contain several public entities in addition to private entities. 

We split each email into chunks of up to $150$ words, resulting in 112k total passages. We deduplicate the emails to prior to generating passage pairs shown to the crowd workers, resulting in the final set of 47k passages. Duplicates exist because the same email can appear in reply chains, forward chains, and in multiple inbox folders (e.g., sent and received email folders).

\paragraph{Further Processing} Since emails can be quite long, we use a sliding window approach to generate documents, given the sequence length limitations of the transformer architecture \cite{alberti2019bertnq}. Finally, we deduplicate the emails for the final private corpus. We release all the code for preprocessing, annotating, filtering, and deduplication along with the benchmark.

We release all preprocessing code for the public and private data.

\subsection{Passage pairs for bridge questions} We need to generate passage pairs for $\mathrm{Hop_1}, \mathrm{Hop_2}$ of two Wikipedia documents (Public, Public), an email and a Wikipedia document (Public, Private and Private, Public), and two emails (Private, Private).

\paragraph{Public-Public Pairs} For Public-Public Pairs, we use a directed Wikipedia Hyperlink Graph, $G$ where a node is a Wikipedia article and an edge ($a$, $b$) represents a hyperlink from the first paragraph of article $a$ to  article $b$. The entity associated with article $b$, is mentioned in article $a$ and described in article $b$, so $b$ forms a \textit{bridge}, or commonality, between the two contexts. 
Crowdworkers are presented the final public document pairs $(a, b) \in G$. We provide the title of $b$ as a hint to the worker, as a potential anchor for the multi-hop question.

To initialize the Wikipedia hyperlink graph, we use the KILT KnowledgeSource resource \cite{petroni-etal-2021-kilt} to identify hyperlinks in each of the Wikipedia passages. \footnote{\url{https://github.com/facebookresearch/KILT}} To collect passages that share enough in common, we eliminate entities $b$ which are too specific or vague, having many plausible correspondences across passages. For example, given $a$ representing a ``company'', it may be challenging to write a question about its connection to the ``business psychology'' doctrine the company ascribes to ($b$ is too specific) or to the ``country'' in which the company is located ($b$ is too general). To determine which Wiki entities to permit for $a$ and $b$ pairings shown to the workers, we ensure that the entities come from a restricted set of entity-categories. The Wikidata knowledge base stores \text{type categories} associated with entities (e.g., ``Barack Obama'' is a ``politician'' and ``lawyer''). We compute the frequency of Wikidata types across the $5.2$ million entities and permit entities containing any type that occurs at least 1000 times.
We also restrict to Wikipedia documents containing a minimum number of sentences and tokens. The intuition for this is that highly specific types entities (e.g., a legal code or scientific fact) and highly general types of entities (e.g. countries) occur less frequently.

\paragraph{Pairs with Private Emails} Unlike Wikipedia, hyperlinks are not readily available for many unstructured data sources including the emails, and the non-Wikipedia data contains both private and public (e.g., Wiki) entities. Thus, we design the following approach to collect passage pairs involving a private passage. 

We first collect entity occurrences in emails:
\begin{enumerate}
    \item To annotate the public and private entity occurrences in the email passages, we collect candidate entities with the SpaCy NER tagger. \footnote{\url{https://spacy.io/}}
    \item  We split the full set into candidate public and candidate private entities by identifying Wikipedia linked entities amongst the spans tagged by the NER model.  We annotate the text with the open-source SpaCy entity-linker, which links the text to entities in the Wiki knowledge base, to collect candidate occurrences of global entities.
    \footnote{\url{https://github.com/egerber/spaCy-entity-linker}} We use heuristic rules to filter remaining noise in the public entity list. 
    \item We post-process the private entity lists to improve precision.  High precision entity-linking is critical for the quality of the benchmark: a query assumed to require the retrieval of private passages $a$ and $b$ should not be unknowingly answerable by public passages. 
    After curating the private entity list, we restrict to candidates which occur at least 5 times in the deduplicated set of passages.
\end{enumerate}

A total of 43.4k unique private entities and 8.8k unique public entities appear in the emails, and 1.6k private and 2.3k public entities occur at least 5 times across passages. We present crowd workers emails containing at least three total entities to ensure there is sufficient information to write the multi-hop question.

Private-Private Pairs are pairs of emails that mention the same private entity $e$. The Private-Public and Public-Private are pairs of emails mentioning public entity $e$ and the Wikipedia passage for $e$. In both cases, we provide the hint that $e$ is a potential anchor for the multi-hop question.

\paragraph{Comparison Questions} For comparison questions, Wikidata types are readily available for public entities, and we use these to present the crowdworker with two passages describing entities of the same type. For private emails, there is no associated knowledge graph so we heuristically assigned types to private entities, by determining whether type strings occurred frequently alongside the entity in emails (e.g., if ``politician'' is frequently mentioned in the emails in which an entity occurs, assign the ``politician'' type).

\subsection{Data Collection Procedure}
\begin{figure*}[t!]
    \centering
    \includegraphics[width=0.55\linewidth]{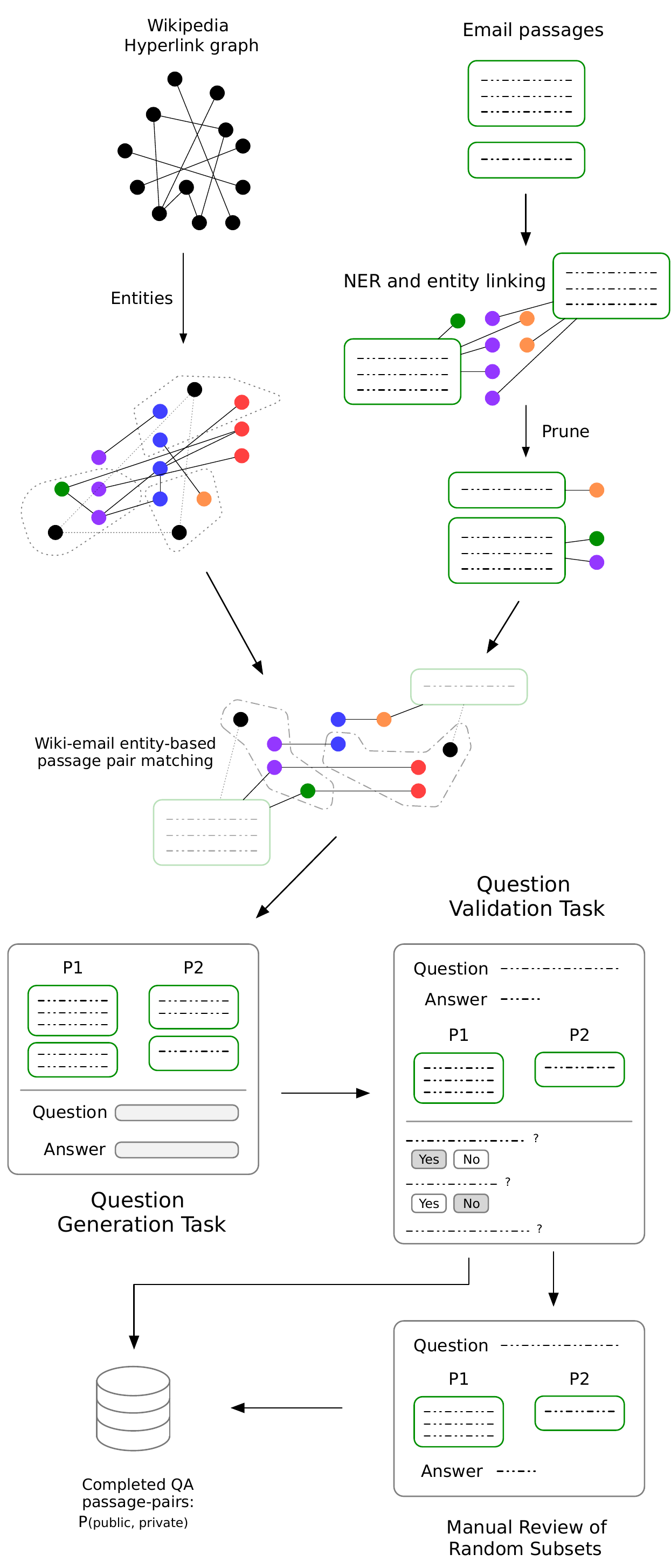}
    \caption[width=\linewidth]{The end-to-end data collection pipeline includes (1) passage-pair generation: we identified entities appearing in emails and Wikipedia passages and categorized entities as public or private, (2) question generation: workers are asked to write a question-answer pair given two passages and a hint stating the common entities in the passages, (3) validation: workers answer a series of questions about each generated question-answer pair to filter out low-quality questions. During the entire process, we manually review questions and workers' understanding of the task.}
    \label{fig:data_collection}
\end{figure*}

Algorithm \ref{CHalgorithm} and Figure \ref{fig:data_collection} give the full data collection procedure for \datasetname. It is adapted from Algorithm 1 in \citet{yang2018hotpotqa}, which was used to produce HotpotQA. 

\begin{algorithm}[t!]
\caption{Data collection procedure}
\label{CHalgorithm}
\begin{algorithmic}[1]
\State \textbf{Input}: Entity-annotated public passages, Entity-annotated private passages, private entity set, public entity set

\While{not finished}
\State $type  =$  random() $<0.2$ // This reflects whether the question is a bridge or comparison question. 
\State $hop_1 =$  random() $<0.5$
\State $hop_2 =$  random() $<0.5$
\newline
\State // Bridge questions
\If{$type$}
\If {$hop_1$ and $hop_2$}
\State \parbox[t]{\dimexpr\linewidth-\algorithmicindent*3}{Uniformly sample public documents $a$, $b$, where entity $e$ appears in both. In the Wikipedia setup, we take $(a, b) \in G$, for hyperlink graph $G$.}
\ElsIf {$hop_1$ and not $hop_2$}
\State \parbox[t]{\dimexpr\linewidth-\algorithmicindent*3}{Uniformly sample a public entity $e$ corresponding to public document $a$. Uniformly sample a private document $b$, which contains $e$.}
\ElsIf {not $hop_1$ and $hop_2$}
\State \parbox[t]{\dimexpr\linewidth-\algorithmicindent*3}{Uniformly sample a public entity $e$ appearing in private document $a$. Uniformly sample a public document $b$, which contains $e$.}
\ElsIf {not $hop_1$ and not $hop_2$}
\State \parbox[t]{\dimexpr\linewidth-\algorithmicindent*3}{Uniformly sample a private entity $e$. Uniformly sample a pair $(a, b)$ where $a$ and $b$ are private documents containing $e$.}
\EndIf\EndIf \newline
\State // Comparison questions
\If{not $type$}
\If {$hop_1$ and $hop_2$}
\State \parbox[t]{\dimexpr\linewidth-\algorithmicindent*3}{Uniformly select a public entity type, and uniformly select two public documents $a$ and $b$ about two different entities $e_a$, $e_b$ of this type.}
\ElsIf {($hop_1$ and not $hop_2$) or (not $hop_1$ and $hop_2$)}
\State \parbox[t]{\dimexpr\linewidth-\algorithmicindent*3}{Uniformly select a private entity type, and uniformly select a public $a$ and private $b$ document about a public and private entity ($e_a$, $e_b$) of this type.}
\ElsIf {not $hop_1$ and not $hop_2$}
\State \parbox[t]{\dimexpr\linewidth-\algorithmicindent*3}{Uniformly select a private entity type, and uniformly select two private documents $a, b$ about two different private entities $e_a$, $e_b$ of this type.}
\EndIf\EndIf

\State \parbox[t]{\dimexpr\linewidth-\algorithmicindent*2}{Workers ask a question about documents $a$ and $b$, given $e$ (or $e_a$, $e_b$ for comparison questions) as an optional anchor.}
\EndWhile
\end{algorithmic}
\end{algorithm}

\begin{figure*}[t!]
    \centering
    \includegraphics[width=0.9\linewidth]{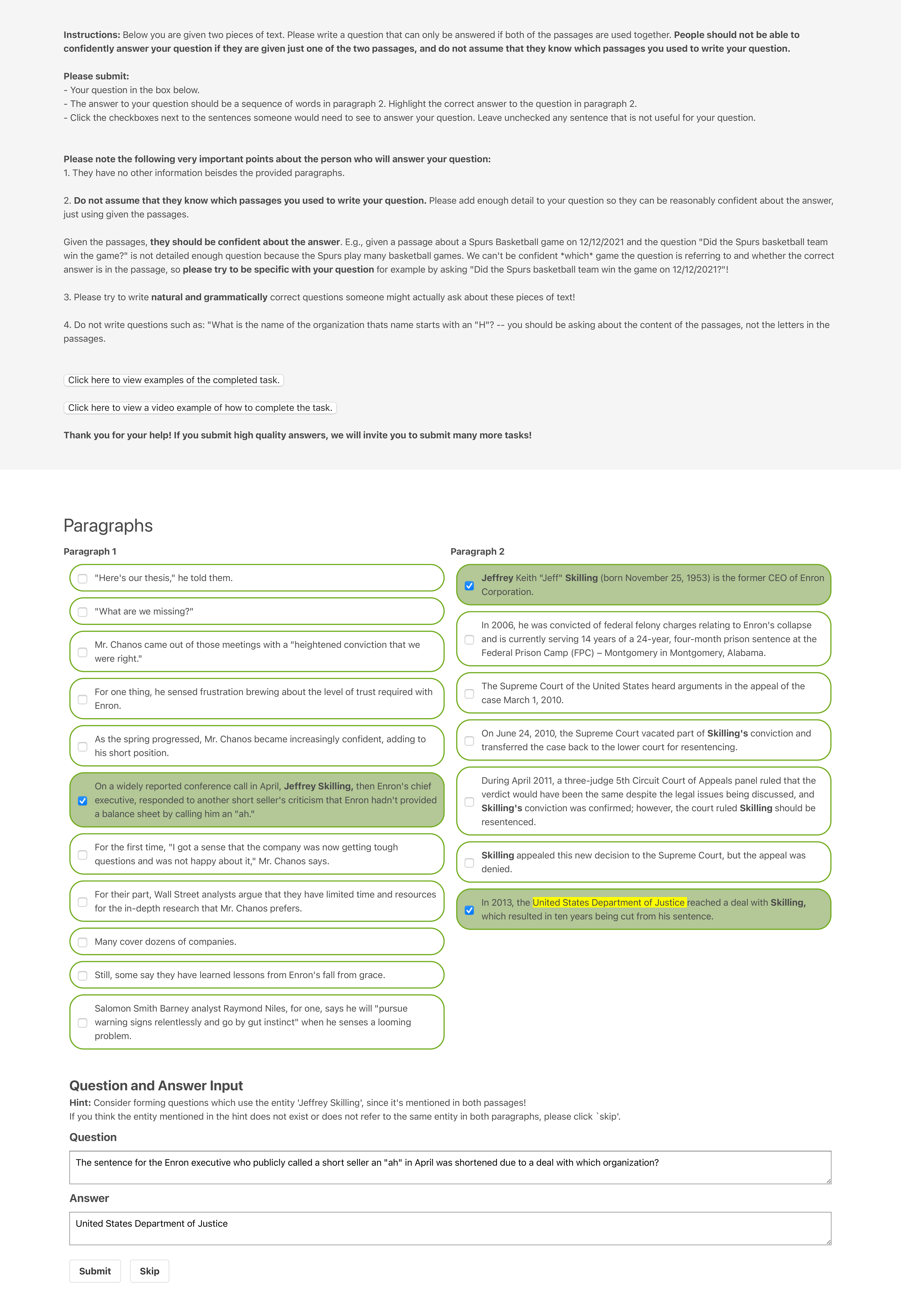}
    \caption[width=\linewidth]{Mechanical Turk interface for \datasetname data collection. Crowdworkers select checkboxes for supporting passages, highlight the answer span, and write the question in the text box.}
    \label{fig:turk_interface}
\end{figure*}

\paragraph{Crowd Worker Interface} 
We use the Mephisto framework to build our crowd worker interface. Figure \ref{fig:turk_interface} gives an example of the interface shown to workers. \footnote{\url{https://github.com/facebookresearch/Mephisto}}

\section{Additional Details for Selective Prediction}

The QA model predicts an answer span in each of the top $k$ passages by predicting the start and end tokens of the answer in each passage. The model outputs scores for each answer and the system outputs the top-scoring answer span. To obtain scores for Section \ref{sec:selpred}, we computed the softmax over all $k$ scores and selected the top score. The same model (trained on the full HopotQA data) was used for all private and non-private runs in Section \ref{sec:selpred}.

Since multiple passages can be used to answer a question, especially as discussed in the case of email-based questions, we also tried identifying groups within the top $k$ answer spans for which the model predicted the same answer --- we then combined softmax scores at the group level and used the top group score as $c$. This resulted in $68.1\%$ coverage at 53.0 F1 (non-private baseline) for \datasetname, but a much lower $71.4\%$ coverage at 75.0 F1 for HotpotQA.






\end{document}